\documentclass[lettersize,journal]{IEEEtran}

\usepackage{amsmath,amsfonts}
\usepackage{algorithmic}
\usepackage{array}
\usepackage[caption=false,font=normalsize,labelfont=sf,textfont=sf]{subfig}
\usepackage{textcomp}
\usepackage{stfloats}
\usepackage{url}
\usepackage{verbatim}
\usepackage{graphicx}
\usepackage{balance}

\usepackage{url}
\usepackage{xcolor}
\usepackage{multirow}

\usepackage{makecell}

\usepackage[commandnameprefix=always,final]{changes}

\usepackage[normalem]{ulem}

\newcommand{\etal}{\textit{et al. }}

\begin{document}

\title{Few-Shot Open-Set Hyperspectral Pixel-Wise Classification with Spatial Aware Label Propagation}

\author{Matheus B. Pereira, %
        Jefersson A. dos Santos,~\IEEEmembership{Senior Member, IEEE}
        and Laurent Najman,~\IEEEmembership{Senior Member, IEEE} %
\thanks{Matheus B. Pereira is with the Department of Computer Science, Universidade Federal de Minas Gerais, Belo Horizonte, Brazil, and with the Laboratoire d’Informatique Gaspard Monge, Université Gustave Eiffel, Noisy-le-Grand, France. (e-mail: matheuspereira@dcc.ufmg.br).}%
\thanks{Jefersson A. dos Santos is with the  Department of Computer Science, University of Sheffield, Sheffield, U.K (e-mail: j.santos@sheffield.ac.uk).}%
\thanks{Laurent Najman is currently a full professor in the Mathematics department at Khalifa University (UAE), on leave from the Laboratoire d’Informatique Gaspard Monge, ESIEE Paris, Université Gustave Eiffel, Noisy-le-Grand, France. (email: laurentalain.najman@ku.ac.ae)}%
\thanks{The authors would like to thank Serrapilheira Institute (grant \#R-2011-37776) and Brazil National Council for Scientific and Technological Development (CNPq). This work was conducted during a scholarship supported by the International Cooperation Program CAPES/COFECUB Fondation at Université Gustave Eiffel. Financed by CAPES – Brazilian Federal Agency for Support and Evaluation of Graduate Education within the Ministry of Education of Brazil.}%
\thanks{Manuscript received April 30, 2026.}} %

\markboth{IEEE Journal of Selected Topics in Applied Earth Observations and Remote Sensing,~Vol.~X, No.~X, Month~202X}%
{Pereira \MakeLowercase{\textit{et al.}}:Few-Shot Open-Set Hyperspectral Pixel-Wise Classification with Spatial Aware Label Propagation}

\maketitle

\begin{abstract}
Pixel-wise classification of hyperspectral images is a fundamental task in remote sensing, but its deployment in real-world applications is often limited by the scarcity of labeled data and the presence of unknown classes at inference time. In this work, we propose \textbf{FOSSeL}, a unified framework for few-shot, open-set hyperspectral pixel-wise classification that does not require supervised training. The proposed approach first leverages self-supervised learning to construct a discriminative feature space from unlabeled data. Then, given a small set of labeled seed pixels, class predictions are obtained through spatially-aware graph-based label propagation over superpixel representations, enabling the joint exploitation of spectral similarity and spatial consistency. To address the open-set scenario, we introduce simple yet effective strategies to identify pixels that do not belong to the known classes. We evaluate FOSSeL on three benchmark hyperspectral datasets under multiple few-shot settings. The results show that the proposed framework achieves competitive or superior performance compared to fully supervised methods while maintaining robustness in the presence of unknown classes. These findings highlight the potential of combining self-supervised feature learning with graph-based inference for practical hyperspectral image analysis in low-annotation and open-world scenarios.
\end{abstract}


\section{Introduction} \label{sec:introduction}

\IEEEPARstart{E}{arth} observation technologies have advanced quickly in recent years, both in terms of modern infrastructure for new sensors/platforms and newly developed data processing tools, which is an important topic in the research community \cite{hosseiny2023beyond}. Remote sensing data plays a crucial role in this context, allowing a comprehensive study of the Earth's surface at different scale levels \cite{khlifi2023graph}, with applications in land use, environmental monitoring, natural resource management, among other tasks \cite{fang2024open}. 

Hyperspectral imaging techniques are essential for several remote sensing applications, allowing for the discrimination of subtle spectral differences \cite{li2019deep}. The result of such techniques is hyperspectral images (HSI), which may contain hundreds of spectral bands that exceed the perception range of the human eye \cite{yue2022spectral}. In this type of imagery, each pixel can be viewed as a high-dimensional vector representing the spectral reflectance of the inspected surface at a specific wavelength \cite{li2019deep}. 

One of the main problems when dealing with hyperspectral data processing is pixel-wise classification, also commonly named in the literature as HSI classification or recognition. In this task, the objective is to classify each pixel in the input image into a land cover class, such as buildings, rivers, roads, and different types of vegetation, among others \cite{yue2022spectral}. This is an important field of research in the hyperspectral community, as proposed methods often face challenges related to the large spatial variability of spectral signatures and limited available training samples \cite{li2019deep}.

Following a similar trend to other computer vision tasks, HSI pixel-wise classification results have been successfully improved in recent years due to advances in deep learning techniques \cite{mdl4ow,li2019deep}. Despite the existence of robust classification methods in the literature, an important problem arises when most of the proposed approaches work under a closed-set scenario in which all training and testing pixels come from the same label space~\cite{oliveira2023fully}. In real-world applications, however, this assumption is often difficult to satisfy, since the learning samples are limited and will hardly represent the whole universe of potential classes to appear in the deployment phase. As a result, the existence of unknown or novel classes will lead to false positives, reducing the precision of the model~\cite{mdl4ow}. 

Open-set pixel-wise classification methods aim at solving the aforementioned problem of identifying pixels that do not belong to the same set of classes seen during training, while also correctly classifying the pixels from the known classes. Although there is a relatively high number of papers in the literature tackling the open-set problem with typical natural images (RGB), only a few approaches are focused specifically on remote sensing data, and even fewer methods are suitable for HSI \cite{openpixel,ssel}. Due to the highly different characteristics of hyperspectral imagery and the common problems that follow this type of data (small amount of available data, fewer classes, and low openness), it is difficult to directly use open-set techniques developed for daily image recognition \cite{mdl4ow}. Thus, in this work, we propose a novel framework for HSI pixel-wise classification considering both the problem of few available data for training (few-shot learning) and the existence of unknown classes during testing (open-set recognition). Our approach, \textbf{F}ew-shot \textbf{O}pen-set hyperspectral recognition from \textbf{S}elf-\textbf{S}up\textbf{e}rvised \textbf{L}earning training (FOSSeL), is a generalizable framework that allows for multiple solutions to be implemented on top of the base of the pipeline that performs closed-set predictions from a few labeled samples, similarly to a semi-supervised method.

\begin{figure*}[ht]
    \centering
    \includegraphics[width=\linewidth]{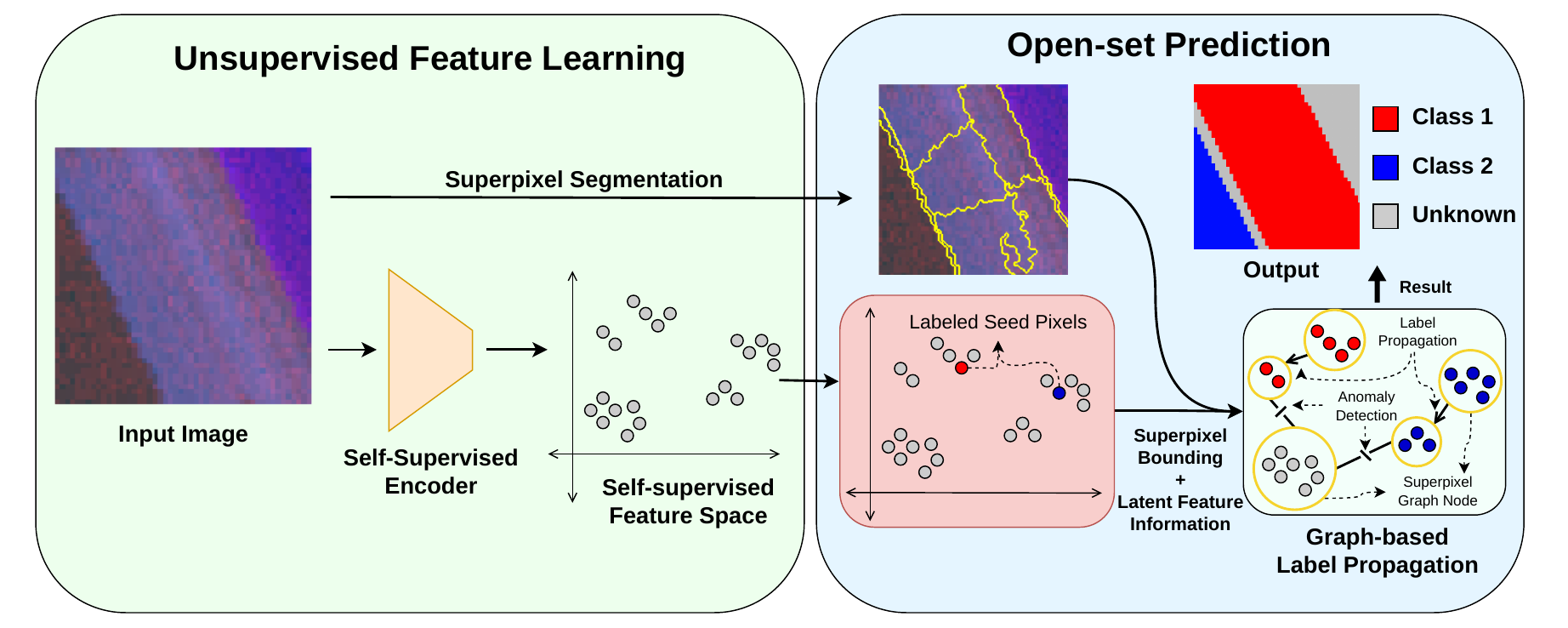}
    \caption{Overview of the proposed FOSSeL framework. Given a feature space obtained through self-supervised training (see Fig. \ref{fig:ssl_overview}) and a few labeled seed pixels, the open-set prediction module, which can represent several possible algorithms, classifies known pixels and identifies unknown classes, generating the final output.}
    \label{fig:overview}
\end{figure*}

We show that FOSSeL is capable of classifying known classes for an input HSI from a few annotated pixels, while also recognizing as unknown the ones that do not belong to the set of classes of the previously labeled samples. This is done in a pipeline that allows for solutions that do not require supervised training, as the feature space is generated from self-supervised (SSL) training and the prediction of the pixels is done via label propagation in a graph constructed on top of a super-pixel segmentation,
which further allows for the exploitation of the spatial relationship between pixels along with the spectral information. Figure \ref{fig:overview} presents an overview of the proposed FOSSeL pipeline.

While recent works have explored unsupervised representations and graph-based propagation independently, a unified approach that explicitly addresses the open-set problem in highly constrained few-shot settings remains unexplored. A comprehensive review of the relevant literature, specifically covering open-set recognition and segmentation for both RGB and HSI data, and semi-supervised learning, is detailed in Section~\ref{sec:related}. Additionally, it is important to note that this study focuses on pixel-wise classification in open-set scenarios. The resolution of sub-pixel fractional abundances (i.e., spectral unmixing of mixed pixels) falls outside the scope of this work.

FOSSeL does not include in its framework the use of the widely employed cross-entropy loss. This is relevant because the performance of methods based on cross-entropy is often limited due to the obtaining of features with low representation, resulting in embedding spaces with weak discrimination capabilities between classes \cite{brilhador2024prototypical}. 
We exploit and adapt graph-based label-propagation algorithms enforcing spatial connectivity as a means of helping the prediction in an open-set scenario. However, the generalizable aspect of the framework also allows for the implementation of solutions that rely on cross-entropy-based strategies if desired. 

We evaluated FOSSeL on three hyperspectral datasets with different characteristics, showing the robustness of the method under varying circumstances and surpassing the results achieved by several state-of-the-art methods from the literature. The code for the proposed framework is publicly available at \url{https://github.com/matheusbarrosp/FOSSeL}.%

The main contributions and innovations of the paper are the following:

\begin{itemize}
    \item \textbf{Flexible Self-Supervised Open-Set Framework:} We propose \textbf{FOSSeL}, a unified and flexible framework for few-shot, open-set hyperspectral pixel-wise classification that does not require supervised training and can be adapted to different feature extractors and label propagation strategies.
    \item \textbf{Unsupervised Feature Validation:} We demonstrate that self-supervised feature spaces are highly effective for open-set hyperspectral classification, even with extremely limited labeled samples.
    \item \textbf{Novel Spatially-Aware Propagation:} We introduce spatially-aware graph-based label propagation strategies that jointly exploit spectral similarity and spatial consistency for robust classification.
\end{itemize}

To validate these contributions, we empirically demonstrate that the proposed framework achieves competitive or superior performance compared to fully supervised state-of-the-art methods across multiple datasets under strict few-shot settings.

Unlike prior approaches that rely on supervised training or fixed architectures, the proposed framework decouples feature learning from pixel classification, enabling the use of self-supervised representations combined with flexible graph-based inference. This design makes FOSSeL particularly suitable for real-world remote sensing scenarios where labeled data are scarce and unknown classes are expected during deployment.

The remainder of the paper is organized as follows. In Section~\ref{sec:related}, we discuss related works in the literature that also tackle the open-set problem in HSI recognition. Section~\ref{sec:methodology} presents the FOSSeL pipeline and describes how it functions. In Section~\ref{sec:experimental}, we detail the experimental setup, presenting the dataset and the baselines used for comparison. Section~\ref{sec:results} presents the results obtained by FOSSeL. Finally, in Section~\ref{sec:conclusion}, we conclude the paper.

\section{Related Works} \label{sec:related}

Hyperspectral image pixel-wise classification, also known as HSI recognition, is a well-studied, although still hot, topic in the remote sensing area.  Mapping land cover classes in satellite imagery is a problem that can be modeled as the supervised classification of each pixel, with the closed-set scenario being the standard assumption. A wide range of solutions has been proposed for this problem in the literature. This includes solutions based on convolutional neural networks \cite{kumar2024binary, wang2021fully}, graph convolutional networks \cite{jamali2024attention, xu2024fuzzy}, recurrent neural networks \cite{zhou2021multiscanning, zhou2023multiscanning}, transformers \cite{scheibenreif2023masked, zhang2023lightweight}, generative adversarial networks \cite{yu2024robust, qin2023distance}, etc. Despite that, none of these methods are originally suited for open-set scenarios, which is a common and important constraint when considering applying the proposed solutions to real-world problems.

Additionally, other problems associated with HSI pixel-wise classification exist and have been previously explored in the literature, such as end-member extraction or spectral unmixing. Mixed pixels (pixels that cover multiple ground materials, for example) bring more challenges to the problem, since they may account for more than one of the classes of the dataset. End-member extraction is the task that aims to select a collection of ``pure'' signature spectra of the materials present in the scene \cite{plaza2012endmember}. Although we acknowledge that taking into consideration the pureness of the pixel when trying to solve the open-set HSI pixel-wise classification problem could be an important step for more robust results, in this work, we do not consider this aspect of the data. Instead, we will focus on the methodology and evaluation based only on the annotation directly available for each dataset.

\subsection{Open-set Recognition and Segmentation}

The open-set problem (OSR) has been explored more extensively for natural RGB image classification. Starting with the formalization in the work of Scheirer \textit{et al.} (2012) \cite{scheirer2012toward}, several methods have been proposed since then. Geng \textit{et al.} (2021) \cite{geng2021survey} separate the work in the OSR literature into discriminative models and generative models. For discriminative ones, OpenMax \cite{openmax} was one of the seminal deep-learning-based works, a method that replaces the traditional softmax layer with a new layer that adds an unknown output class to the existing $K$ known categories. From there, different approaches were explored in the literature, usually resorting to statistical modeling of data or its features \cite{nunes2023systematic} and often estimating an anomaly score and/or employing Extreme Value Theory (EVT) strategies \cite{openmax, oza2019c2ae, yoshihashi2019classification, yang2022open}. Another trend observed in the literature, reconstruction-based methods \cite{oza2019c2ae, sun2020conditional} leverage the reconstruction error of auto-encoders to differentiate known samples from unknown ones \cite{nunes2023systematic}. In terms of generative models, approaches like G-OpenMax \cite{g-openmax} and Neal \textit{et al.} (2018) \cite{Neal_2018_ECCV} aim to generate open-set samples with the use of generative adversarial networks (GANs) as a means of helping the model learn what is different from the known training samples.

Open-set semantic segmentation (OSS) is a more difficult problem due to its pixel-wise approach. Some papers propose solutions for non-hyperspectral remote sensing data \cite{openpcs, openpixel, coreseg, hong2022goss, brilhador2024prototypical}, which share characteristics with HSI, such as high background complexity, class imbalance, and tiny foreground objects \cite{li2021cvpr}. However, unlike HSI, rich hyperspectral information is unavailable for this type of data, and the imagery usually presents much higher spatial resolutions. Therefore, despite it being possible to adapt the architecture of most of these methods to work with hyperspectral inputs, the fact that the specificity of this type of data is not being exploited may lead to suboptimal results, which is a motivation to seek solutions for HSI.

\subsection{Open-set Hyperspectral Image Pixel-wise Classification}

Recently, the open-set problem for HSI has started to catch the attention of the community. MDL4OW \cite{mdl4ow} was one of the first methods proposed in the context of open-set HSI pixel-wise classification. Following a similar trend to previous reconstruction-based methods \cite{oza2019c2ae, coreseg}, MDL4OW simultaneously conducts classification and reconstruction of the input data, where the reconstruction output acts as a tool to help separate unknown classes. Another method following a similar strategy, SSLR \cite{sslr} employs spatial and spectral latent reconstruction to find unknown classes, classifying unknown samples with the help of a defined threshold.

OCN \cite{ocn} avoids the need for manual thresholding by following a meta-learning-based strategy. OCN learns to reject outliers by optimizing a cross-entropy loss in meta-training episodes where a subset of known classes is treated as pseudo-unknowns. SSMLP-RPL \cite{ssmlp-rpl} is a method that introduces reciprocal point learning (RPL) in the open-set context. Working on feature space learning, RPL aims to push the known classes to the periphery of the space according to reciprocal points in the training phase, with unknown samples composing the extra-category space.

OSC-SCL \cite{osc-scl} is a method that employs contrastive learning to constrain spectral-spatial features, so that features of the same class can be tightly clustered in the feature space. Unknown classes are found with a threshold based on the distance to the class anchors.
FCPN \cite{fcpn} defines prototypes from the extracted features, and a cosine similarity is used to classify unknown objects. EVML \cite{evml} is a method that tries to predict a manually annotated ``coarse class'' for unknown regions. For each known land cover class, EVML fits a Weibull distribution based on the support feature distance from the respective prototypes.

HyperCASR \cite{hypercasr} is a reconstruction-based method that does not focus on the reconstruction of the original pixels, but instead works on extracted semantic features with the aim of reducing the impact of noise from the input data. Finally, more recently, CACL \cite{cacl} employs transformer-based methods to model the spatial position information of input tokens. A contrastive technique is used to bring samples from the same class together, while pushing away those from different classes. Unknown samples in the testing phase are predicted with an adaptive segmentation threshold.

Our approach is different from the aforementioned works due to several characteristics. Firstly, FOSSeL is a highly generalizable framework, allowing for different implementations of solutions on top of self-supervised training, which is an advantage when compared to methods that present a fixed architecture. Our framework allows for solutions that do not require any type of supervised learning, which can be useful in scenarios where the available annotated samples are highly limited. Finally, our approach efficiently exploits the spatial information of the input data due to the use of graph-based label propagation techniques, while previous works generally rely only on deep learning-based strategies for leveraging spatial information.

\subsection{Semi-supervised Learning}

Semi-supervised learning is another paradigm that is closely related to the way we approach the open-set HSI problem. Semi-supervised learning aims to improve learning performance with a few labeled data inputs by using additional unlabeled samples, exploring the latent patterns of the unlabeled data \cite{gui2024survey}. A few papers have tackled the open-set recognition problem from a semi-supervised perspective.

In Ma \textit{et al.} (2023) \cite{inl}, the authors propose INL, a model that leverages unlabeled data in the framework by training a rotation prediction branch and classifying unknown samples by checking their distances to the learned class prototypes. SSB \cite{ssb} is a method that incorporates pseudo-labels with high inlier classifier confidence into the training. The authors train the model with pseudo-labels from the inlier classifier whose confidence scores are above a predefined threshold, while the outlier detector consists of one-vs-all binary classifiers. Se-FOSS \cite{se-foss} is a method that utilizes a self-supervised consistency loss in an open-set scenario, leveraging all types of unlabeled data and employing an energy-based score to recognize data belonging to known classes.

FOSSeL can be considered similar to semi-supervised learning approaches due to leveraging both unlabeled data and the annotation information of a few labeled samples. However, FOSSeL allows for the implementation of solutions that do not include any type of training with the labeled information from the few annotated pixels, with the only training occurring without labels with a self-supervised strategy. Another point is that semi-supervised methods often employ pseudo-label strategies\cite{gui2024survey}. Due to its generalizable properties, FOSSeL can be adapted to employ this type of strategy in the framework, but this is not required. In essence, the similarities between FOSSeL and semi-supervised learning come from the similar nature of self-supervised and semi-supervised learning, with more direct implementations of semi-supervised strategies being possible to be implemented due to the open nature of the framework.

\section{Methodology} \label{sec:methodology}

As illustrated in Figure~\ref{fig:overview}, FOSSeL employs a two-stage pipeline whose final result is the classification of all input pixels from the HSI, including the separation of the ones that are unknown compared to the few available annotated samples. The first stage is a self-supervised training that aims to generate a distributed feature space across the classes of the dataset. The second stage involves classifying the pixels in the input image using a graph constructed over a superpixel segmentation. This can be done without further training by exploiting the spatial relationships among superpixels to propagate the labels of known samples to the remaining unlabeled ones. 

\subsection{Self-Supervised Feature Generation} \label{sec:methodology_ssl}

Deep-based supervised learning often requires a high amount of labeled data for training~\cite{jaiswal2021survey}. Remote sensing tasks are usually scarce in terms of annotation due to the nature of the applications, which may require the work of a specialist to correctly perform the labeling process. Thus, self-supervised learning (SSL) methods emerge as a promising solution, as they are capable of avoiding the need for fully annotated large-scale datasets by adopting self-defined pseudo-labels as supervision \cite{jaiswal2021survey}. In the context of hyperspectral data, the use of SSL has been explored recently \cite{wang2022sslrs} to the point of showing the capability to outperform traditional supervised learning \cite{braham2022self}.

The first step of the proposed FOSSeL framework is to perform SSL training with all the available data of the target dataset without the need for any labels. This process has the objective of generating a well-distributed feature space for all the pixels of the input HSI, where a classification algorithm can later perform the prediction. While the FOSSeL framework is modular and can accommodate various SSL architectures, in this work we experimented with the feature extraction module using VICReg \cite{vicreg}. The theoretical justification for this choice is twofold. First, unlike contrastive SSL methods (such as SimCLR\cite{Chen2020simclr} and MoCo\cite{He2020moco}) that heavily rely on mining negative pairs on large batch sizes, VICReg prevents representation collapse purely through information maximization by an explicit constraint on the variance of the embeddings, which was shown to play the role of a negative term between the vectors \cite{vicreg}. In the context of HSI, defining true negative pairs is usually risky due to low spatial resolution and high inter-class similarity. Furthermore, VICReg explicitly regularizes the variance and covariance of the embeddings. The covariance regularization forces the learned variables to be decorrelated, which is particularly beneficial for hyperspectral data where spectral bands are typically highly correlated. This approach makes the method robust even with low-dimensional projectors, offering a good balance between discriminative representation for few-shot scenarios and computational efficiency \cite{vicreg}, serving as a robust baseline to evaluate our proposed label propagation mechanics. Other works in the literature, such as Braham \etal (2022) \cite{braham2022self} with a Barlow-Twins method \cite{barlowtwins}, have also achieved results that outperform supervised learning before. 

Following VICReg's methodology, let $X \in \mathbb{R}^{H \times W \times C}$ be a hyperspectral image with height $H$, width $W$, and $C$ spectral bands. At every SSL training iteration, we sample a patch $p$ of size $s \times s$ from the $X$, i.e., $p \in \mathbb{R}^{s \times s \times C}$. In traditional training, this patch would be used to generate two different views of the same input, however, in this work, we follow the pair sampling approach proposed by Braham \etal (2022) \cite{braham2022self} in which another random patch with at least $50\%$ overlap to the first one, $p' \in \mathbb{R}^{s \times s \times C}$, is also selected. The idea is that neighbor-overlapping patches should have similar representations, thus serving as a type of data augmentation that exploits the spatial locality of the patches, introducing a spatial regularity prior to the SSL training \cite{braham2022self}. Then, two transformations, $t$ and $t'$, are sampled from a distribution $T$ in order to generate different views of $p$ and $p'$: $x = t(p)$ and $x'=t'(p')$. These views are fed to a shared backbone encoder $f$ that will generate their latent representations $r=f(x)$ and $r'=f(x')$, which will later be used as the final feature for each pixel of the HSI for classification. During the SSL training, these representations are mapped to a projection head $h$ that expands the representations to embeddings, $z=h(r)$ and $z'=h(r')$, where the loss is computed. VICReg's loss function is a weighted average of the invariance, variance, and covariance terms (more details can be found in VICReg's original paper \cite{vicreg}), defined as:

\begin{equation}
    L(Z,Z') = \lambda s(Z, Z') + \mu [v(Z) + v(Z')] + \nu [c(Z) + c(Z')],
    \label{eq:ssl_loss}
\end{equation}
where $\lambda$, $\mu$ and $\nu$ are hyper-parameters that define the importance of each term, $s(.)$, $v(.)$ and $c(.)$ represent, respectively, the invariance, variance and covariance regularization terms, while $Z = [z_1, ..., z_n]$ and $Z'=[z'_1, ..., z'_n]$ represent batches of embeddings generated by the projection head for $n$ different patches from the HSI.

The transformations applied to the input image are an important step for the training of an SSL model. For FOSSeL, along with the aforementioned pair sampling operation that selects a random neighbor overlapping patch to form a pair with the first randomly chosen one, spatial and spectral transforms are applied to the input images. Considering operations that work in the spatial domain, we employed random horizontal flipping, random vertical flipping, and random rotation by multiples of 90 degrees. As for transformations that work in the spectral domain, we apply Gaussian noise, random band dropping, and random band swapping between adjacent bands, inspired by operations that were previously successfully used in the literature \cite{braham2022self}.

\begin{figure*}[ht]
    \centering
    \includegraphics[width=\linewidth]{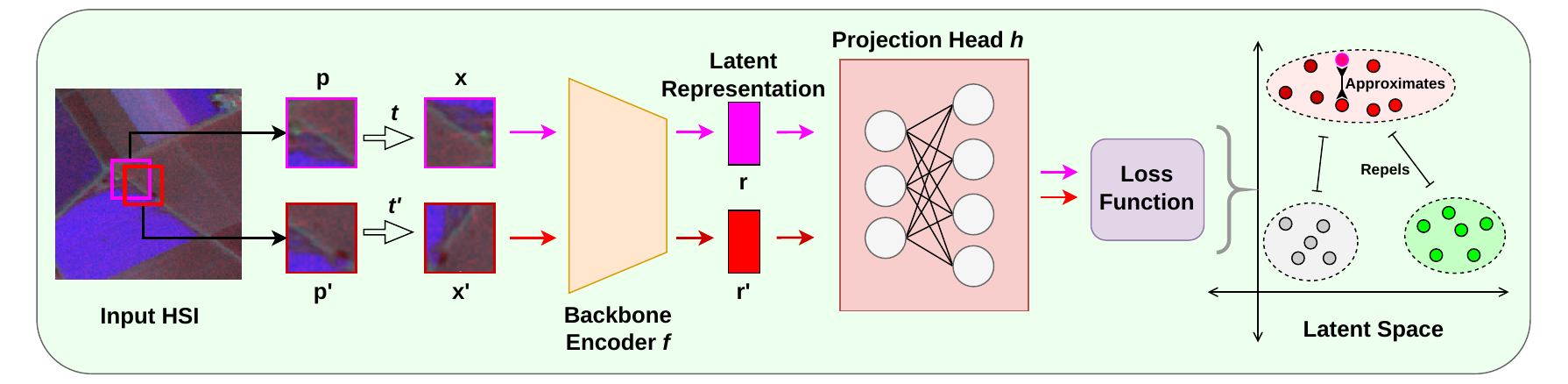}
    \caption{Overview of the SSL training process. A random patch $p$ and its overlapping neighbor $p'$ are transformed and processed by the backbone encoder to generate the desired representation. During training, the projection head maps the representations to embeddings where the loss function is calculated, which allows for similar samples to be brought together in the feature space, separating them from the dissimilar ones.}
    \label{fig:ssl_overview}
\end{figure*}

The backbone encoder $f$ of the SSL model can be designed from scratch or adapted from the many neural network architectures that can be found in the literature. Our implementation of FOSSeL employs as its backbone encoder a simple ResNet18 \cite{resnet} for its lightweight architecture and reduced number of parameters necessary for training. As for the projection head $h$, we employ an MLP-like architecture, comprised of a sequence of two fully connected layers followed by a $1D$ batch normalization layer and ReLU activation, with a last fully connected layer generating the final embedding. Figure~\ref{fig:ssl_overview} illustrates the entire SSL training process.

\subsection{Pixel Classification}

Once the SSL pre-training is complete, the projection head $h$ is discarded, as the remainder of the framework works with the latent representation learned by the backbone encoder $f$. The idea is that a well-trained encoder will be able to generate a feature space where all the classes are separated and distributed, thus later facilitating the work of the classifier. 

For all of the following methods, we consider that a few labeled pixels are provided as seeds to the algorithm. From these seeds, the label will be propagated to the remaining unlabeled pixels according to the chosen algorithm. To prevent ambiguity regarding the role of labeled data in our framework, it is crucial to distinguish between the feature generation and pixel classification phases. The feature generation phase of FOSSeL is strictly unsupervised: the network weights are optimized via self-supervised learning entirely without the use of ground-truth labels. Once this phase is complete, the network parameters are permanently frozen. During the subsequent pixel classification phase, supervision information is indeed introduced in the form of a very limited number of labeled seed pixels. However, in this phase, the provided labels act exclusively as starting anchors for the graph propagation algorithms. Thus, the seeds are never used to compute a loss gradient, fine-tune the feature space, or update the model parameters. Therefore, while supervision information is required to define the known classes during inference, the underlying representation learning of the framework remains completely unsupervised.

In the literature, previous works have turned their attention to graph-based frameworks, which allow for addressing the high dimensionality and the scarcity of labeled samples problems in HSI \cite{challa2019watersheds, tripletwatershed}. These approaches have shown promising results by constructing a graph that connects the pixels based on spectral similarity and/or spatial adjacency. For example, the Watershed Classifier \cite{challa2019watersheds} has been shown to work well with HSI \cite{tripletwatershed} by exploiting connectivity patterns and spatial relationships without the need for trainable parameters. Furthermore, a graph-based structure is well-suited for label propagation algorithms, which propagate the label information from a few known seeds (known pixels) to the entire set of pixels to be classified, thus achieving robust classification performance even with minimal training data.

However, working with graphs where each pixel represents a vertex can be computationally demanding, as the number of vertices does not scale well for increasingly bigger inputs. Thus, we employ superpixel segmentation as a preprocessing step for the pixel prediction step (not used during the SSL training), as shown in Figure~\ref{fig:overview}. This approach reduces the number of vertices in the graph, as each superpixel can be seen as a collection of neighboring pixels that are perceptually and/or spectrally similar.

\begin{figure*}[ht]
    \centering
    \includegraphics[width=\linewidth]{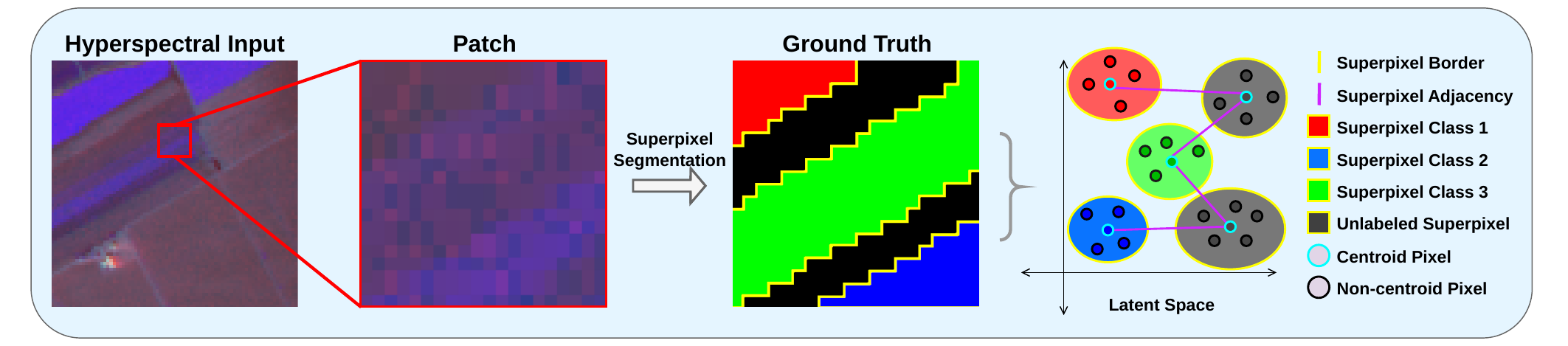}
    \caption{Illustration of how the graph is constructed from the input HSI and how the pixels are organized in the learned SSL feature space. Each computed superpixel is a vertex and connects to its adjacent neighbors. The centroid pixel of each superpixel is chosen as its representative, being used to compute the edges' weights according to the Euclidean distance in the latent space of centroids to adjacent superpixels.}
    \label{fig:graph_creation}
\end{figure*}

In the proposed FOSSeL framework, each superpixel represents a vertex in the graph, while the edge set is defined by the superpixels' adjacencies. Thus, each superpixel (vertex) is connected to its adjacent neighbors, with the weight of the edge being defined by the Euclidean distance in the SSL feature space of the centroid pixel (chosen as the ``representative'' pixel of the superpixel). This way, adjacent vertices, represented by superpixels' centroids, that belong to different classes would ideally be positioned far from each other in the feature space, setting a high value for the edge's weight, while vertices neighboring other superpixels from the same class would be positioned close in the feature space, resulting in a low edge weight. As the SSL training does not use labeled data, these feature distances are learned based on the inherent spectral information of the pixels. Figure~\ref{fig:graph_creation} illustrates an example of how this process works. With this approach, the computational cost of the label propagation algorithm is reduced, enabling faster execution times. On top of the computational advantage, the use of superpixels intrinsically incorporates spatial information into the framework. Instead of operating on individual pixels, label propagation occurs between spatially coherent regions, introducing a form of regularization that makes the framework more robust to noise, ensuring that the propagated labels are consistent with object structures and the local spatial context.

As the proposed FOSSeL framework is generalizable and allows for different strategies of pixel (or superpixel) classification to be employed, we selected three different approaches to systematically evaluate the performance of the method with varying graph-based techniques. These methods were selected to represent different paradigms for exploiting the spatio-spectral information of the pixels that is encoded in the graph. The first is a pathfinding algorithm, which classifies a node based on the distance to the closest labeled seed. The second approach is a label propagation model similar to the classic approach proposed by Zhu and Ghahramani (2002) \cite{labelpropagation}, where labels are propagated from a row-stochastic transition matrix constructed on top of the graph's adjacencies. Finally, the last method propagates the labels with a k-Nearest Neighbors (k-NN) approach, considering the representations in the SSL-trained feature space.

One important aspect that is common for all proposed strategies is that for the pixels inside the same superpixel region as a seed, the framework always sets their labels to the same as the respective labeled seed, regardless of the prediction method, as we consider the similarity of these pixels in the superpixel space a stronger evidence of a shared class than any prediction the algorithm may make.

\subsubsection{Closest Seed Propagation Method (CSProp)}

The CSProp method operates on a graph $G=(V,E,W)$ where $V$ represents the set of vertices (superpixels), $E$ represents the set of edges, and $W:E \to \mathbb{R}^+$ represents the weight assigned to each edge (Euclidean distance of centroids' SSL representation).

This algorithm implements a ``Nearest Seed'' propagation strategy. First, the set of vertices $V$ is partitioned into two groups: a small subset of seed vertices $V_{seed}$, for which class labels are known, and the set of unlabeled ``test'' vertices $V_{test}$.

The objective of the method is to find the closest seed for every test vertex. To achieve this, a multi-source Dijkstra algorithm is executed on the graph, originating from all seed vertices in $V_{seed}$. This process calculates, for every test vertex $v_j \in V_{test}$, the shortest path to its nearest seed vertex $v_i \in V_{seed}$. The label propagation is then performed directly: each unlabeled vertex $v_j$ is assigned the known class label of its corresponding nearest seed $v_i$. Finally, these newly assigned superpixel labels are mapped back to the original image space, resulting in a thematic map where all pixels within a superpixel region share the same predicted class.

\subsubsection{Similarity Matrix Propagation Method (SimProp)}

This method operates on the principle of information diffusion, aiming to propagate label information from the seed nodes to the remaining ones through the weighted graph, similarly to the method proposed by Zhu and Ghahramani (2002) \cite{labelpropagation}. First, we convert the weights (previously defined as the Euclidean distance in the feature space between two adjacent nodes) to a similarity weight using the following inverse relationship:

\begin{equation}
    w'(e) = \frac{1}{1 + w(e)},
\end{equation}
where $e$ represents an edge $e \in E$ and $w(e)$ represents its previous weight.

Next, the method builds a similarity graph that is represented as an $N \times N$ weighted adjacency matrix $W$, where $N$ is the total number of vertices and the entries $W_{ij}$ correspond to the similarity $w'(e)$ for the edge between nodes $i$ and $j$. This matrix is normalized to create a row-stochastic transition matrix $T$, dividing each row of $W$ by its sum:

\begin{equation}
    T_{ij} = \frac{W_{ij}}{\sum_{k=1}^{N} W_{ik}}
\end{equation}

We also construct a label probability matrix $P$ of shape $N \times C$, where $C$ is the number of known classes. This matrix is initialized as follows:

\begin{itemize}
    \item Seed Nodes: If a node $i$ is a seed of class $k$, its corresponding row in $Y$ is initialized with the value $1.0$ at the $k$-th position and zero in all other positions (one-hot vector);
    \item Unlabeled Nodes: All other nodes are initialized with a vector composed of only zeros.
\end{itemize}

From the transition matrix $T$ and the probability matrix $P$, the algorithm computes the label propagation. For this, every node gathers label information from its neighbors as the weighted average of its neighbors' current label distributions. This operation is done via a single matrix multiplication:
\begin{equation}
    P^{(t+1)} = T \times P^{(t)}
\end{equation}

After the propagation step, to ensure the seed nodes do not lose their ground-truth information, their label distributions are reset to their original one-hot vectors. This process continues iteratively until the difference between the label distributions from two subsequent iterations falls below a tolerance threshold, or until a maximum number of iterations is reached. The final prediction is set as:
\begin{equation}
    pred(v) = \underset{k \in C}{argmax} (P_{v,k})
\end{equation}

\subsubsection{k-NN Propagation Method (KNNProp)}

Unlike the previous methods, the k-nearest neighbors (k-NN) approach aims to construct a graph where the connections are based on the neighborhood of a node in the feature space, rather than taking into account the original superpixel adjacencies from the input image.  For this, an edge is created between a node $i$ and a node $j$ if and only if node $j$ is one of the $k$ nearest neighbors of node $i$ in the feature space.

Once the k-NN graph is built, the algorithm iteratively propagates label information similarly to the previous SimProp method via transition matrices. Thus, the main difference between the two approaches lies in the graph itself, more specifically, in the adjacencies that are considered.

\subsection{Opening the Closed-set Predictions}

After executing any of the previously presented label propagation methods, a closed-set classification of the dataset is generated as a result. This can be enough for some applications that are only concerned about recognizing known classes. For the final part of the framework, we explore how to open this closed-set prediction to separate the unknown classes from the rest.

We are interested in performing an open-set classification of the HSI, which means that there are pixels from unknown objects or classes that should not be mislabeled as one of the known annotated classes. Thus, the last question to be answered is: how do we separate the unknown pixels from the ones that belong to the same class as the available annotated seeds? For each one of the three presented methods to propagate the seed labels, we propose an alternative or modification of the algorithm to determine which pixels belong to unknown classes. It is important to note, however, that the previously explained process, from the SSL pre-training to the closed-set prediction, allows for several possibilities to identify unknown classes to be experimented with as well.

One important remark about FOSSeL is that for the proposed approaches, no label is ever provided for unknown pixels, i.e., the user only needs to provide a few seeds for the known classes. It is possible to adapt the method to also take into consideration that the user may label/provide seeds for unknown classes, however, in this work, we only take into consideration the scenario in which the users are interested only in defining what they want to be recognized in the image, rather than having to point towards what they do not want to be recognized as well.

\subsubsection{Opening CSProp Method}

To detect unknown nodes in CSProp, we employ a thresholding technique according to the shortest distances to the seeds. From the closed-set prediction of the previous step, we have access to the value of the distance to the closest seed of node $v$:
\begin{equation}
    d_{\text{seed}}(v) = min_{s \in S} d(v, s),
\end{equation}

With the shortest distance calculated for each vertex, we have access to a global distribution of seed distances for all nodes in the graph, $D_{\text{seed}} = \{d_{\text{seed}}(v) \mid \forall v \in G\}$. We then define the OOD threshold, $d_{\text{mean}}$, as the \textbf{mean} of this entire distribution:

\begin{equation}
    d_{\text{mean}} = \text{mean}(D_{\text{seed}})
\end{equation}

Finally, any node $v$ is classified as unknown if its seed distance exceeds this mean threshold.

\subsubsection{Opening SimProp and KNNProp Methods}

Both SimProp and KNNProp employ a similar label propagation algorithm. Their main distinction lies in the graph construction: SimProp connects nodes based on spatial superpixel adjacencies, while KNNProp connects them based on k-nearest neighbors in the feature space. Thus, the same strategy can be employed to open these two methods.

For these methods, we leverage the closed-set label probability distribution as a measure of classification confidence. Let $Y$ be the final $N \times C$ label probability matrix, where $N$ is the number of nodes and $C$ is the number of classes. For any node $v$, its probability distribution is the vector $Y_v = [p_1, p_2, \dots, p_C]$. We define the confidence score for this node, $\text{conf}(v)$, as the maximum probability in its distribution:

\begin{equation}
    \text{conf}(v) = \max(Y_v)
\end{equation}

We then establish a minimum confidence threshold, $\tau$. Any node $v$ whose final confidence score is lower than this threshold is classified as unknown.

\begin{equation}
    pred\_open(v) = \begin{cases} 
      argmax (Y_v) & \text{if } \text{conf}(v) \geq \tau, \\
      \text{unknown} & \text{if } \text{conf}(v) < \tau
    \end{cases}
\end{equation}

The proposed thresholding strategies are intentionally simple and do not require additional training or calibration. Empirically, we observed that these criteria remain stable across datasets and few-shot settings, providing a good balance between correctly classifying known samples and identifying unknown ones. While more sophisticated thresholding strategies could be employed, the adopted approach prioritizes robustness and ease of deployment in practical scenarios.

\subsection{Methodological Novelty}

While the FOSSeL framework builds upon foundational concepts in self-supervised learning \cite{vicreg} and graph-based semi-supervised label propagation \cite{labelpropagation}, it is important to emphasize that our approach is not a trivial concatenation of these methods. Standard SSL methods are not inherently designed for open-set anomaly isolation, and traditional label propagation algorithms frequently fail when applied to the noisy, high-dimensional space of hyperspectral imagery without spatial regularization. The main innovation of FOSSeL lies in formulating a unified framework designed for the few-shot, open-set hyperspectral domain.

\section{Experimental Setup} \label{sec:experimental} 

\subsection{Datasets}

We evaluated the FOSSeL framework with three hyperspectral datasets from the literature with different characteristics: Indian Pines, Kennedy Space Center, and Salinas. For the evaluation of our methods in an open-set scenario, we follow a similar protocol to other works in the literature \cite{yue2022spectral, mdl4ow, oliveira2023fully, brilhador2024prototypical} to adapt datasets that were originally made for closed-set cases. Thus, for each dataset, a subset of classes is selected to be hidden and treated as unknown. This simulates the situation where the remaining classes of the dataset present at least one labeled pixel, therefore being known, while the unknown ones have no label available for training. During the evaluation, we ignore, for all datasets, the pixels that have no label in the ground truth, as we can not determine if they are unknown or not due to the possibility that they contain samples from otherwise known classes that were simply not annotated.

\subsubsection{Indian Pines (IP)}

This dataset was gathered by the Airborne Visible/Infrared Imaging Spectrometer (AVIRIS \cite{aviris}) sensor over Northwestern Indiana. The scene presents 200 spectral reflectance bands over $145 \times 145$ pixels, with a wavelength range of $0.4-2.5 \times 10^{-6}$ meters.

The dataset contains 16 labeled classes: alfalfa, corn no-till, corn min till, corn, trees, pasture, pasture mowed, hay-windrowed, oats, soybeans no-till, soybeans min till, soybeans clean till, wheat, woods, buildings grass trees, and stone-steel towers. From these, three were selected to be hidden and treated as unknown: alfalfa, pasture mowed, and oats. For the choice of the hidden classes, we followed the same set selected in Pal \textit{et al.} (2022) \cite{ocn}.

\subsubsection{Kennedy Space Center (KSC)}

This dataset was also gathered by the AVIRIS \cite{aviris} over the Kennedy Space Center (KSC), Florida, with the same wavelength as IP and from an altitude of approximately 20 km. The scene contains 176 reflectance bands over $512 \times 614$ pixels.

KSC contains 13 labeled classes: scrub, willow swamp, CP hammock, slash pine,  oak/broadleaf, hardwood, swamp, graminoid marsh, spartina marsh, cattail marsh, salt marsh, mud flats, and water. For this dataset, the classes selected to be hidden were scrub, willow swamp, and swamp. The choice for these classes was made by separating the swamp-related classes from the rest, while scrub was additionally selected by randomly picking from the set of classes without a close semantic relation with the others (unlike, for example, the different types of marsh classes).

\subsubsection{Salinas (SA)}

The Salinas dataset is another one that was gathered by the AVIRIS \cite{aviris} over Salinas Valley, California. With 204 hyperspectral bands, its spatial resolution is 3.7-meter pixels in a scene of size $512 \times 217$.

This dataset contains 16 labeled classes: broccoli green weeds 1, broccoli green weeds 2,  fallow, fallow rough plow, fallow smooth, stubble, celery, grapes untrained, soil vineyard develop, corn senesced green weeds, lettuce romaine 4wk, lettuce romaine 5wk, lettuce romaine 6wk, lettuce romaine 7wk, vineyard untrained, and vineyard vertical trellis. The hidden ones for this scene were the lettuce-related classes, which are lettuce romaine 4wk,  lettuce romaine 5wk, lettuce romaine 6wk, and lettuce romaine 7wk. The choice was also based on the set selected by Pal \textit{et al.} (2022) \cite{ocn}, with the inclusion of the lettuce romaine 5wk, which is set for training in their original GitHub code.

\begin{figure*}[!ht]
\centering
\subfloat[]{\includegraphics[width=0.3\linewidth, height=0.3\linewidth]{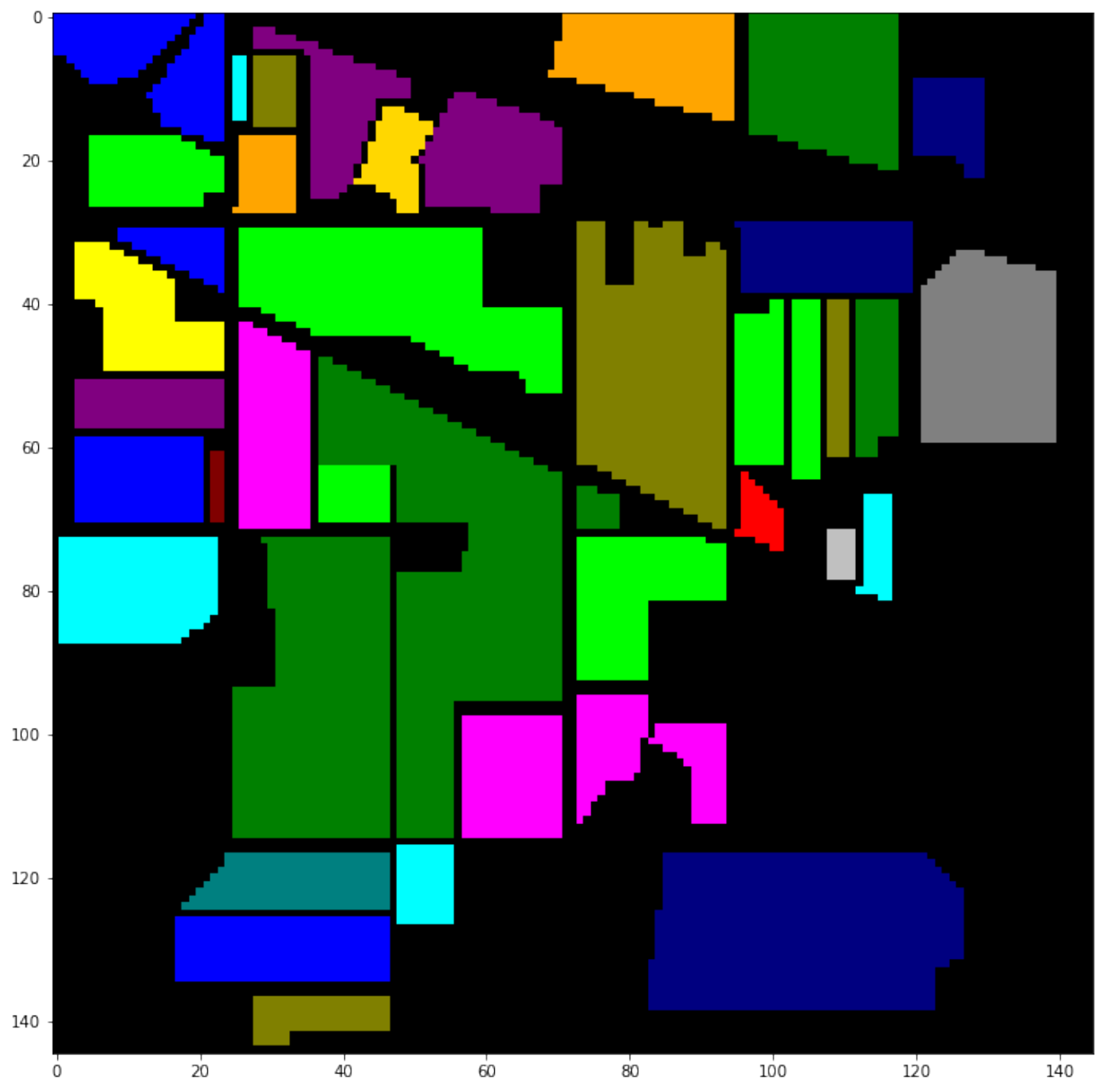}%
\label{fig:ip_label}}
\subfloat[]{\includegraphics[width=0.3\linewidth, height=0.3\linewidth]{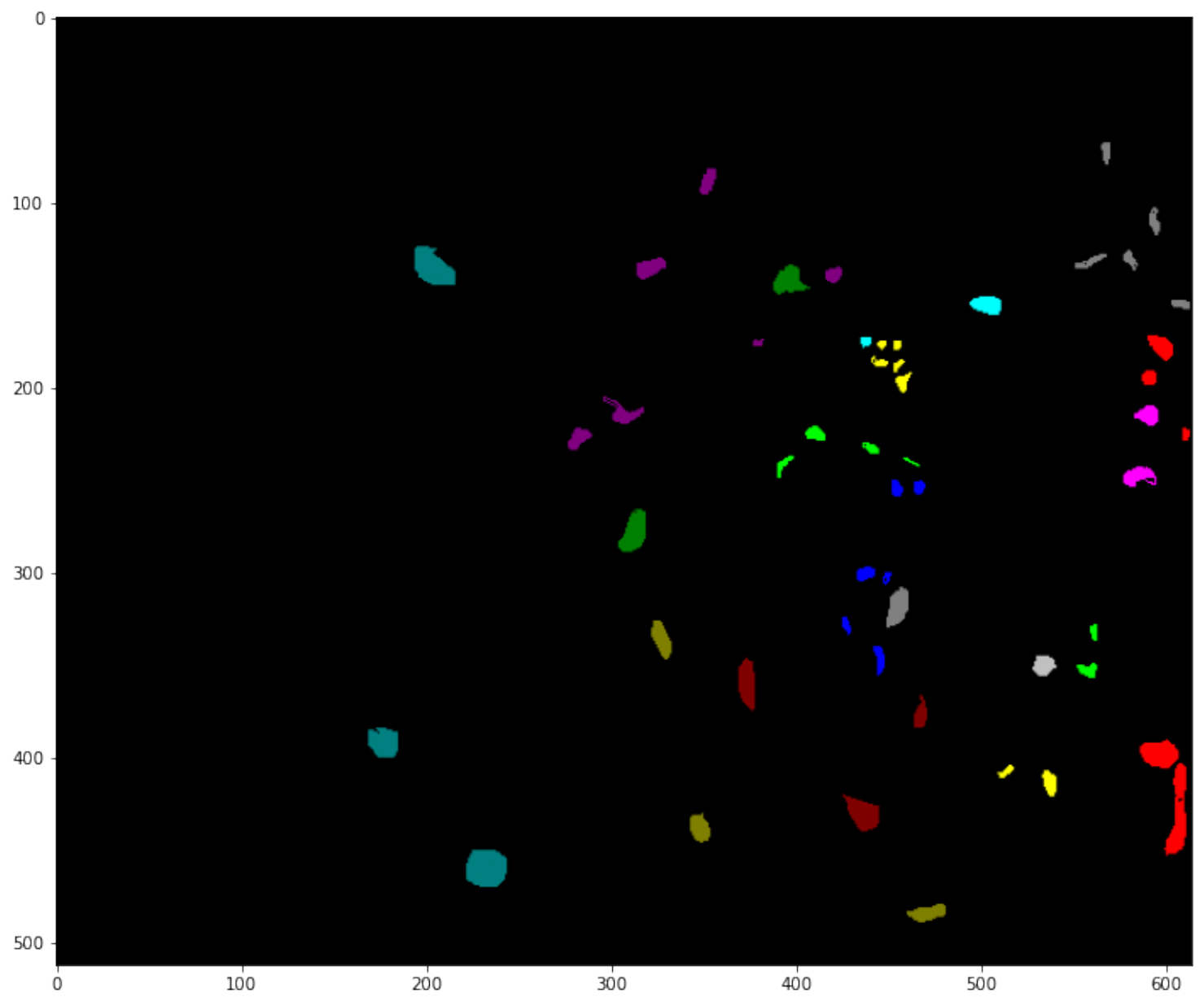}%
\label{fig:ksc_label}}
\subfloat[]{\includegraphics[width=0.3\linewidth, height=0.3\linewidth, keepaspectratio]{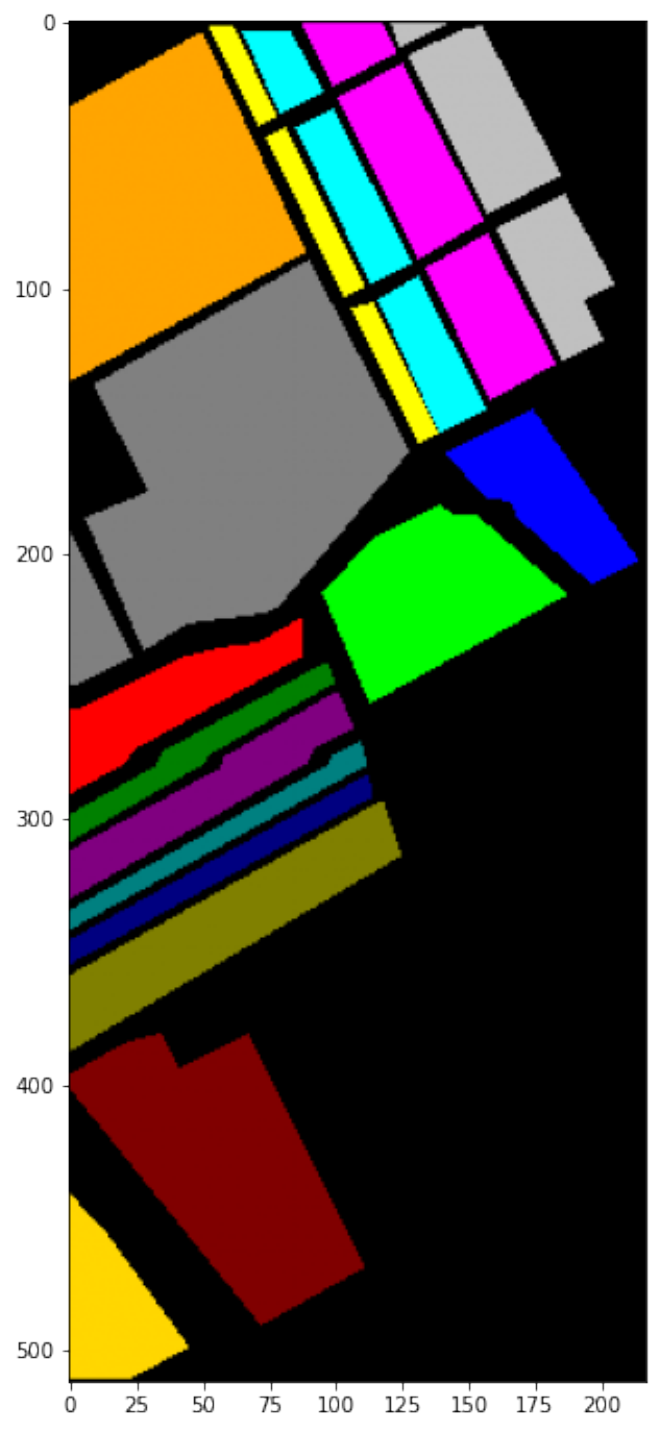}%
\label{fig:sa_label}}
\caption{Visualization of the thematic map of the three datasets. (a) Indian Pines. (b) Kennedy Space Center. (c) Salinas.}
\label{fig:dataset_labels}
\end{figure*}

Figure \ref{fig:dataset_labels} shows the thematic map of each dataset, illustrating the entire set of labeled pixels for the three scenes and coloring in black the pixels that were not labeled (thus not participating in the metrics' calculation). As the figure shows, the datasets present different spatial characteristics beyond the different sets of labeled classes previously detailed. These varying characteristics help to show how well FOSSeL can operate in different scenarios.

\subsection{Implementation Details}

FOSSeL allows for different models to be employed in its framework, including different architectures for the encoder backbone of the SSL training and different methods of open-set classification at the final stage. For our experiments, we used the PyTorch framework \cite{pytorch}, and the specification of our implementation of the methods is as follows.

We employed VICReg \cite{vicreg} as our core model for the self-supervised training of the framework with the Lightly API\footnote{\url{https://github.com/lightly-ai/lightly}}. For the SSL loss presented in Equation \ref{eq:ssl_loss} we set the hyperparameters $\lambda$, $\mu$ and $\nu$ as $25.0$, $25.0$ and $1.0$, respectively. Regarding the backbone encoder architecture, ResNet18 \cite{resnet}, we adapted PyTorch's torchvision implementation by changing the number of input channels of the first layer to the number of hyperspectral bands of the respective dataset. Thus, the encoder starts by performing a dimensionality reduction on the data by reducing the number of channels to $64$ in the first layer. From there, each block in the ResNet18 architecture doubles the number of filters being learned up to $512$, which is the output of the encoder, both for the projection head and for the pixel classification in the second stage of the framework. Finally, the projection head, which follows the MLP-like architecture described in Section \ref{sec:methodology_ssl}, increases the number of features to 2048, which is the final size of the embeddings where the loss is computed.

All three datasets were trained for $20000$ epochs with a batch size of $4096$ on one GeForce GTX Titan Xp GPU with 12GB of memory. We experimented with different patch sizes and learning rate values for each one of the datasets, and the best results were achieved with the following configurations: for IP, patch size $11$ and learning rate $0.04$; for KSC, patch size $9$ and learning rate $0.01$; and lastly, for SA, patch size $11$ and learning rate $0.1$. The optimizer used in the training was the Stochastic Gradient Descent.

The rationality of the selected learning rates and patch sizes was based on the stability of the loss convergence along with the standard deviation of the $\ell_2$-normalized output ($\ell_2 std = z/||z||_2$, where $z$ is the latent embedding vector), as defined by Chen and He (2021) \cite{simsiam} for siamese-based neural network architectures, as tracking this standard deviation provides a reliable metric to prevent representation collapse. We also visually evaluated the discriminative quality of the resulting feature space by inspecting the distribution of the features according to a t-Distributed Stochastic Neighbor Embedding (t-SNE) \cite{t-sne} plot. We focused our experiments on the IP dataset initially, as it is the smallest among the three. Thus, for this dataset, we experimented with learning rate values varying between 0.01 and 0.1 and with patch sizes varying between 3, 5, 7, 9 and 11. For the smallest patch size (3) with 0.1 learning rate, the training ended with a loss value oscillating around $12.86$ and $\ell_2 std$ of  $0.037$ (approaching the theoretical optimum of $1/\sqrt{d} \approx 0.044$, where $d$ represents the dimensionality of $z$.). Other learning rate configurations for this patch size either failed to converge (such as with a 0.5 learning rate, which caused \textit{nan} values) or exhibited slow convergence (such as 0.01 and 0.05 learning rates, which stopped with an $\ell_2 std$ value of $0.034$). In contrast, utilizing the largest patch size (11) with a $0.04$ learning rate yielded a loss of approximately $16.59$, while maintaining an equivalent $\ell_2 std$ of $0.037$. Although the total loss was higher for the largest patch, this is an expected outcome, as the expanded spatial context makes it harder for VICReg's loss invariance term to force augmented views of the same patch into identical representations. Because the $ \ell_2 std$ indicated comparable representation quality across the best cases for each patch size, the final selection of size 11 was driven by the t-SNE feature visualizations (as seen in Figure~\ref{fig:ip_features}), since the increased spatial context provided by the larger patch size generated more compact and well-separated feature clusters.

Building upon the baseline established by IP, we evaluated the SA dataset by testing variations of the patch size of 9, 11 and 13, alongside learning rates of $0.01$, $0.05$ and $0.1$. For this dataset, a $0.1$ learning rate achieved the best results, yielding a mean $\ell_2 std$ of $0.037$ (against the $0.044$ theoretical optimum) across the varying patch sizes, whereas the configurations with the lowest learning rate ($0.01$) achieved a mean $\ell_2 std$ of $0.035$. The best result was ultimately achieved using a patch size of 11, resulting in an $\ell_2 std$ of $0.0375$ and a loss value of $14.11$. As for the KSC dataset, learning rates exceeding $0.05$ caused training instability and poor convergence. With the lowest experimented learning rate of $0.01$, similar results were observed for patch sizes 9, 11 and 13, in which all of the three achieved $\ell_2 std$ of $0.035$. Reducing the patch size to 7 showed a decrease in performance, with $\ell_2 std$ of $0.033$. Given that KSC possesses the highest spatial resolution among the datasets, the final choice between patches 9, 11 and 13 prioritized the option with the more efficient configuration in terms of memory consumption, i.e., patch size 9. Finally, we also conducted additional ablation experiments aimed at increasing the network's focus on the central pixel of the patches. This included omitting the overlapping patch augmentation strategy and artificially assigning a higher weight to the central pixel prior to the forward pass. However, both of these strategies proved detrimental, as they impaired model convergence and resulted in cluttered feature spaces. Table~\ref{tab:ablation_ssl} shows the results of these experiments, with the variations being applied to the baseline configuration of the IP dataset (patch size 11 and 0.04 learning rate).

\begin{table}[h]
    \centering
    \caption{Results of the ablation study of SSL training strategies on the IP dataset. ``Overlap'' denotes the application of the overlapping neighbor augmentation, while ``Central Pixel Weight'' indicates the scaling factor applied to the central pixel of the patch to evaluate the impact of spatial biasing.}
    \label{tab:ablation_ssl}
    \begin{tabular}{cccccc}
        \hline
        \makecell{Learning \\ Rate} & Patch Size & \makecell{Overlap} & \makecell{Central Pixel \\ Weight} & Loss & $\ell_2 std$ \\
        \hline
        0.04 & 11 & yes & none & 16.56 & 0.037 \\
        0.04 & 11 & no & none & 4.86 & 0.036 \\
        0.04 & 11 & no & 2.0 & 4.95 & 0.036 \\
        0.04 & 11 & no & 3.0 & 3.15 & 0.033 \\
        0.04 & 11 & no & 4.0 & 3.19 & 0.032 \\
        \hline
    \end{tabular}
\end{table}

As the results in Table~\ref{tab:ablation_ssl} show, when the overlapping neighbor augmentation strategy is removed (``Overlap: no''), the training loss drops significantly from $16.56$ to $4.86$. While a lower loss traditionally indicates better convergence, in this context, by applying transformations to the exact same patch rather than overlapping neighbors, the network can easily minimize the loss without learning robust spatial-spectral representations. Conversely, the overlapping strategy introduces an important challenge to the training that forces the network to reconcile spatial shifts, which ultimately yielded a more organized and separated feature space, as seen in Figure~\ref{fig:ip_features} along with the higher $\ell_2 std$ value from the table. Furthermore, the results show that artificially increasing the weight of the central pixel did not enhance the overall feature distribution. Between not using an additional weight to the central pixel (``Central Pixel Weight: none'') and using a factor of 2.0, the $\ell_2 std$ was not considerably affected, while further increases in the factor made the representation deteriorate. By heavily biasing the central pixel, the network effectively ignores the surrounding spatial context and reduces the objective to a simple pixel-to-pixel mapping, which prevents the encoder from capturing important spatial-spectral topologies required by the framework for the following label propagation. Therefore, the results confirm that treating all pixels within the patch equally, in conjunction with the overlapping augmentation strategy, allows for the learning of the more organized representations needed for an accurate open-set pixel-wise classification.

For the superpixel segmentation, we used the SLIC (Simple Linear Iterative Clustering) method \cite{slic}. As SLIC is designed for 3-channel color spaces, we first applied Principal Component Analysis (PCA) to the  HSI input, reducing it to its three most significant components. This 3-component image was then used as the input for the SLIC segmentation. For all experiments, we configured the SLIC algorithm to generate approximately 1000 superpixels, except for the experiments on the KSC dataset, where we generated approximately 5000 superpixels due to the higher spatial resolution of the input image. Section \ref{subsec:sensitivity} presents an analysis regarding the impact of the number of segments on FOSSeL's results. A compactness parameter of 15 was used in all experiments.

Finally, regarding the confidence score threshold $\tau$ utilized by \textit{SimProp} and \textit{KNNProp}, its optimal strictness is related to the availability of labeled seeds. To empirically validate this thresholding strategy and demonstrate the adaptable trade-off between known class accuracy and unknown class rejection, we present threshold trade-off curves in Section~\ref{subsec:sensitivity}. As detailed in this subsequent analysis, while extreme few-shot scenarios operate optimally near a permissive $\tau \in [0.5, 0.6]$, scenarios with more available data work better with stricter thresholds to preserve open-set integrity. Thus, we set the threshold as $0.6$ for the scenarios with 5 and 10 seeds, while for the scenarios with 15 seeds we set the threshold as $0.7$.

\subsection{Baselines}

To evaluate the performance of our proposed FOSSeL framework, we establish comparisons against two different categories of baseline methods. The first category consists of a graph label propagation algorithm from unsupervised training based on the Watershed Classifier \cite{challa2019watersheds} and the superpixel-guided label propagation (SGLP) method from DSCA-Net \cite{sglp}. The Watershed Classifier shares our core methodological paradigm, relying on unsupervised training prior to graph-based label propagation, while the SGLP was designed to propagate known labels via a superpixel-wise similarity measure. Thus, both methods serve as direct comparative baselines to validate the efficacy of our propagation strategies. The second category consists of state-of-the-art HSI pixel-wise classification algorithms that operate in a supervised manner. Including these supervised models allows us to evaluate how closely our unsupervised feature space can compete with methods that have the distinct advantage of actively adapting their representations using the labeled data.

\subsubsection{Graph-based Unsupervised Propagation Method}

For the first category, we compare our results to the Watershed Classifier \cite{challa2019watersheds} and the SGLP \cite{sglp} method. Following the definition of Challa \etal (2022) \cite{challa2019watersheds}, the Watershed Classifier works on an edge-weighted graph defined in the same way as in the CSProp method. The algorithm assumes a dissimilarity measure $\rho(x,y)$ between two vertices $x,y \in V$ called \textit{pass value}, which is defined as:

\begin{equation}
    \rho(x,y) = \underset{\pi \in \Pi(x,y)}{min} \text{ } \underset{e \in \pi}{max}W(e),
\end{equation}
where $\pi$ represents a path between the vertices $x$ and $y$ and $\Pi$ represents the set of all possible paths between them. Thus, the pass value represents the minimum of the highest weights across the possible paths from $x$ to $y$. 

Once the pass value is defined, by taking the information of a few labeled pixels (seeds) for each known class of the dataset, one can perform a closed-set classification by propagating the label of the seeds to the remaining unlabeled pixels. To do this, for each unlabeled pixel in the set $V_u \subset V$, we calculate the pass value to all the labeled pixels $V_l \subset V$ (where $V_l \cap V_u = \emptyset$) and set the target unlabeled pixel $v \in V_u$ class to the same class as the seed with the minimum pass value:

\begin{equation}
    pred(v) = class(\underset{l \in V_l}{argmin} \text{ } \rho(v,l)),
    \label{eq:pred_closed}
\end{equation}
where $class(.)$ represents a function that returns the label of the respective vertex.

In order to open the Watershed Classifier results, we can treat each seed point as unknown if its class does not exist. Hence, given a set of $k$ known classes $C = \{1,2,...,k\}$, for each class $i \in C$ we calculate the minimum pass value from the seeds of $i$, $V_l^i$, to the seeds of all the other classes $j \in C | j \neq i$, $V_l^j$:

\begin{equation}
   p_i =  \underset{x \in V_l^i, y \in V_l^j}{min} \text{ } \rho(x,y),
\end{equation}
where $p_i$ is the minimum pass value for the class $i$ to all other known classes $j$.

From the $p_i$ calculated for each class, for a test pixel $v \in V_u$ to be classified as known for class $i$, its pass value that defines its closed prediction to be class $i$ is expected to be smaller than $p_i$. Thus, if it is higher, we classify this pixel as unknown:

\begin{equation}
    pred\_open(v) = 
    \begin{cases}
        pred(v) & \text{if } \underset{l \in V_l}{min} \text{ } \rho(v,l) < p_i \\
        \text{unknown} & \text{otherwise}
    \end{cases},
\end{equation}
where $pred(v)$ is the closed-set prediction defined in Equation~\ref{eq:pred_closed}.

As for SGLP \cite{sglp}, this method was designed to mitigate the negative impact of imbalanced data distribution by leveraging spatial-spectral homogeneity. The process begins by segmenting the input HSI into a set of non-overlapping superpixels. From a few labeled pixels, the unlabeled superpixels are classified according to a superpixel-level nearest neighbor algorithm. For this, the method calculates the cosine similarity between the average spectral signature of the unlabeled superpixel and the mean spectrum of each class of labeled samples, subsequently assigning the label of the most similar class to the entire superpixel. In the original paper \cite{sglp}, SGLP is employed to enable the equal selection of unlabeled samples from each category, aiming to generate a balanced training set for subsequent closed-set training. In this work, however, we focus on using this method as an open-set label propagation strategy to compare the results with FOSSeL's variants. Thus, to adapt SGLP's strategy into the open-set scenario, a confidence-based decision boundary is introduced on top of the cosine similarity propagation: the maximum similarity score is extracted for each target superpixel, which is then evaluated against a dynamically calculated threshold defined as the overall mean of the maximum similarity scores across all superpixels. If a specific superpixel's highest similarity fails to exceed this threshold, the closed-set prediction is rejected, being classified as an unknown region.

\subsubsection{HSI Pixel-wise Classification Supervised Methods}

To evaluate the proposed FOSSeL framework, we select four recent, state-of-the-art baseline methods specifically designed for open-set hyperspectral image pixel-wise classification. These baselines comprise a variety of recent architectures and learning paradigms:
\begin{itemize}
    \item \textbf{OSC-SCL} \cite{osc-scl}: A method utilizing supervised contrastive learning designed for HSI open-set scenarios.
    \item \textbf{SSMLP-RPL} \cite{ssmlp-rpl}: A spectral-spatial MLP-like network that employs reciprocal points learning to bound known classes for HSI data.
    \item \textbf{HyperCASR} \cite{hypercasr}: A recent spectral-spatial HSI open-set recognition model based on category-aware semantic reconstruction.
    \item \textbf{CACL} \cite{cacl}: A state-of-the-art class-aware consistency learning framework specifically designed for open-set semi-supervised hyperspectral image pixel-wise classification.
\end{itemize}

All the tests were conducted using the authors' original GitHub code for each method, with the only modifications being related to specific characteristics of the experiment, such as defining a new set of hidden unknown classes and the number of available training samples, along with adding support for new datasets that were not originally included.

For OSC-SCL \cite{osc-scl}, we set the $\eta$ hyperparameter to $3.5$ for the IP and KSC datasets, while for SA, we set it to $2.0$. The remaining hyperparameters were set to the same default values defined by the authors. However, it is important to note that, for the KSC dataset, the method was not capable of learning valid anchors for some of the known classes, as it did not correctly predict any sample. Thus, in order to avoid invalid anchors, we used the logits of all available samples for the classes that fell into this situation, instead of just the ones that were classified correctly.

For HyperCARS \cite{hypercasr}, we set the same hyperparameter values as suggested by the authors on the dataset that was also evaluated in their work, IP. For KSC and SA, the patch size was set to 19, and the number of Groups in the GPE Module to 4. 

Finally, regarding SSMLP-RPL \cite{ssmlp-rpl} and CACL \cite{cacl}, we used the default hyperparameter values proposed by the respective authors, with the only differences being the aforementioned set of unknown classes and the number of training samples.

\section{Results and Discussion} \label{sec:results}

This section details the empirical results of our experiments. We analyze the results in four separate parts. First, we present the performance of the self-supervised learning (SSL) stage, which serves as the feature extraction backbone. Second, we compare the open-set HSI pixel-wise classification results of FOSSeL with the Watershed Classifier \cite{challa2019watersheds}, which is another method operating on the same methodological base. Third, we evaluate and discuss the results compared to state-of-the-art supervised methods for HSI open-set pixel-wise classification. Lastly, we evaluate the sensitivity of our framework regarding the main hyperparameters previously introduced, such as the number of superpixel segments and the confidence threshold $\tau$.

Overall, the results demonstrate that FOSSeL achieves consistent performance across datasets with different characteristics, maintaining strong accuracy even in extremely few-shot scenarios. Notably, the framework remains competitive with supervised approaches despite not relying on labeled data for training, highlighting the effectiveness of the learned feature space and the proposed propagation strategies.

\subsection{Feature Space Learning} \label{subsec:ssl_results}

\begin{figure}[ht!]
\centering
\subfloat[]{\includegraphics[width=0.95\linewidth]{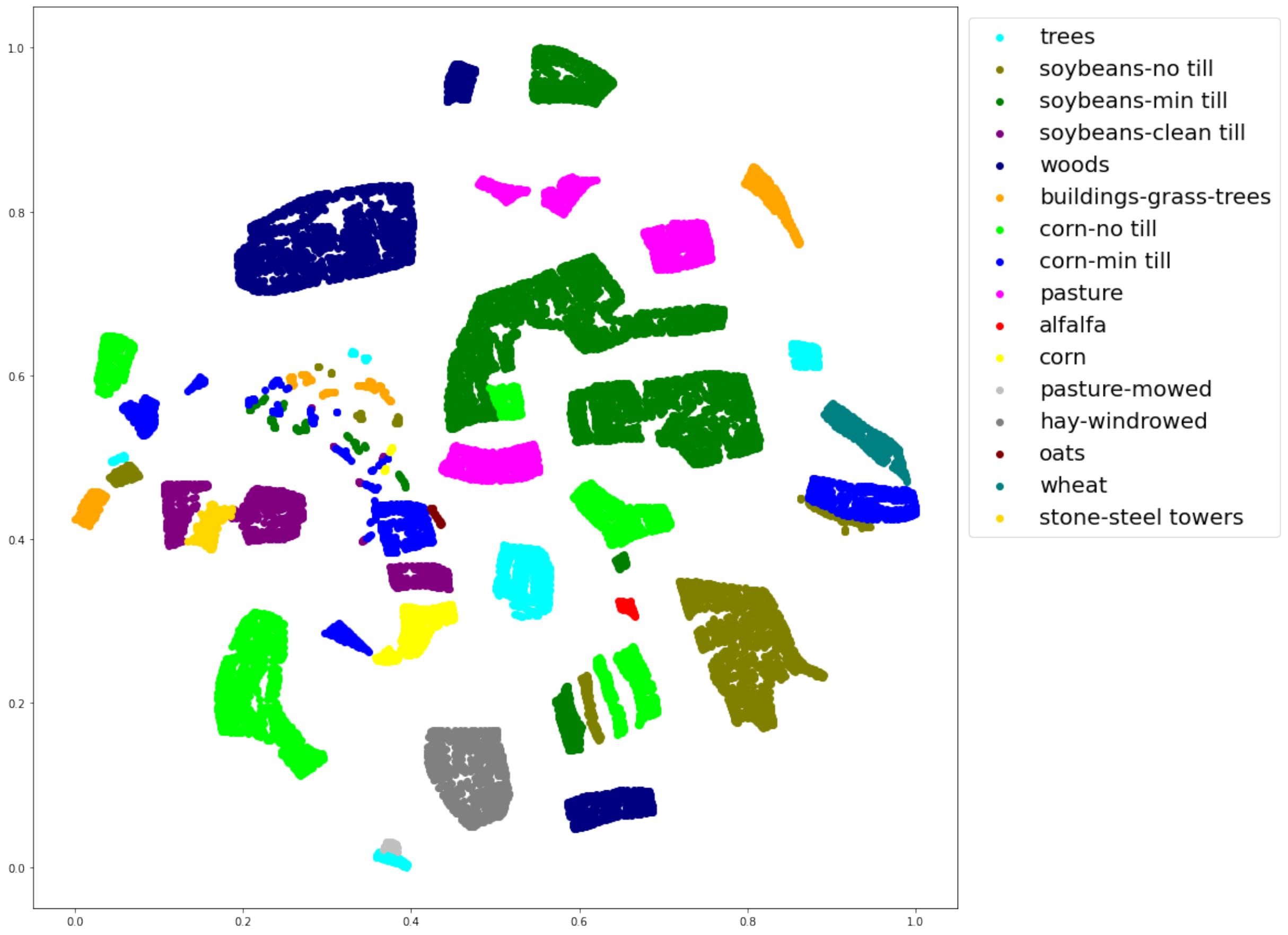}%
\label{fig:ip_features}}
\\
\subfloat[]{\includegraphics[width=0.95\linewidth]{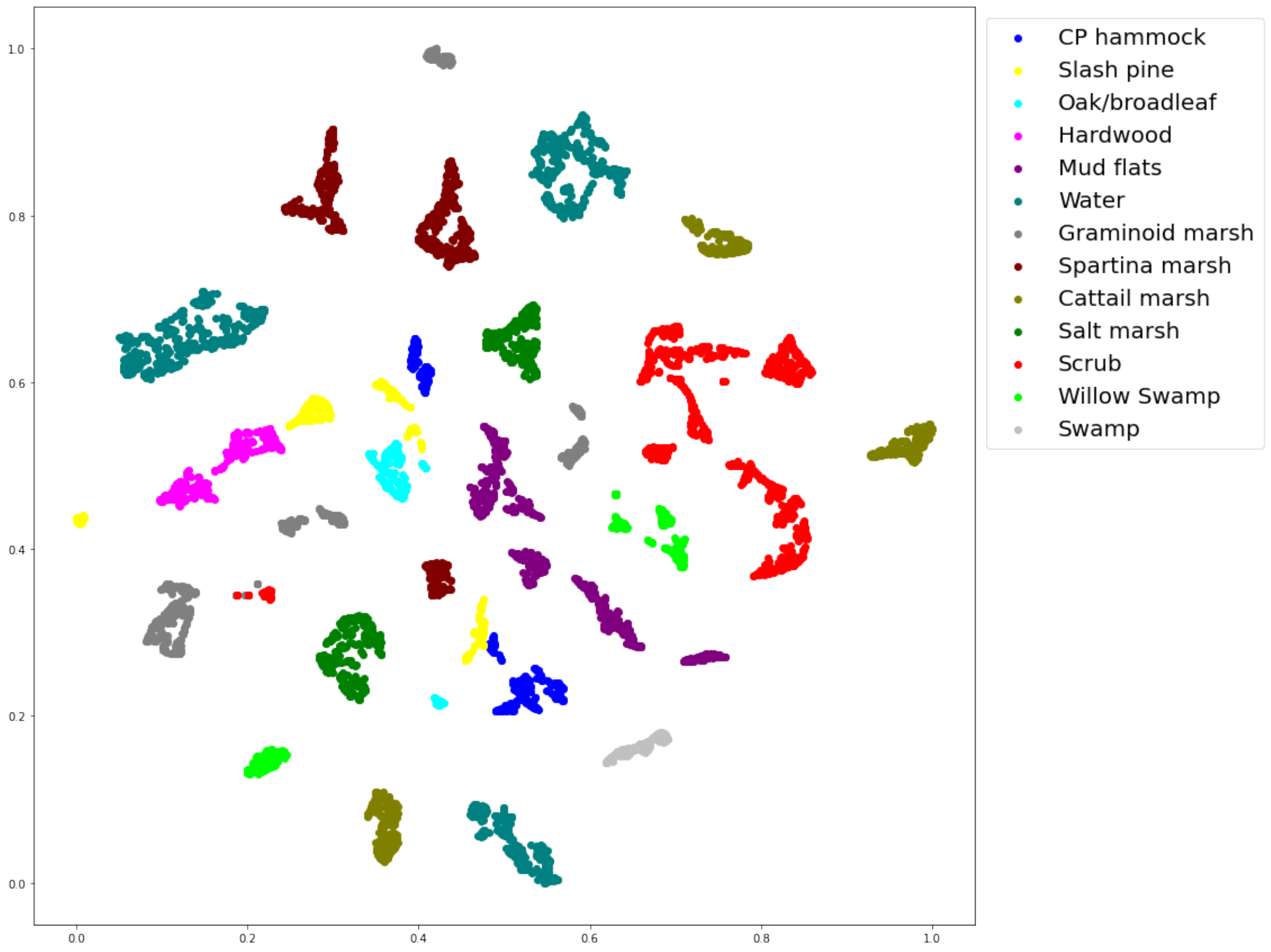}%
\label{fig:ksc_features}}
\\
\subfloat[]{\includegraphics[width=0.95\linewidth]{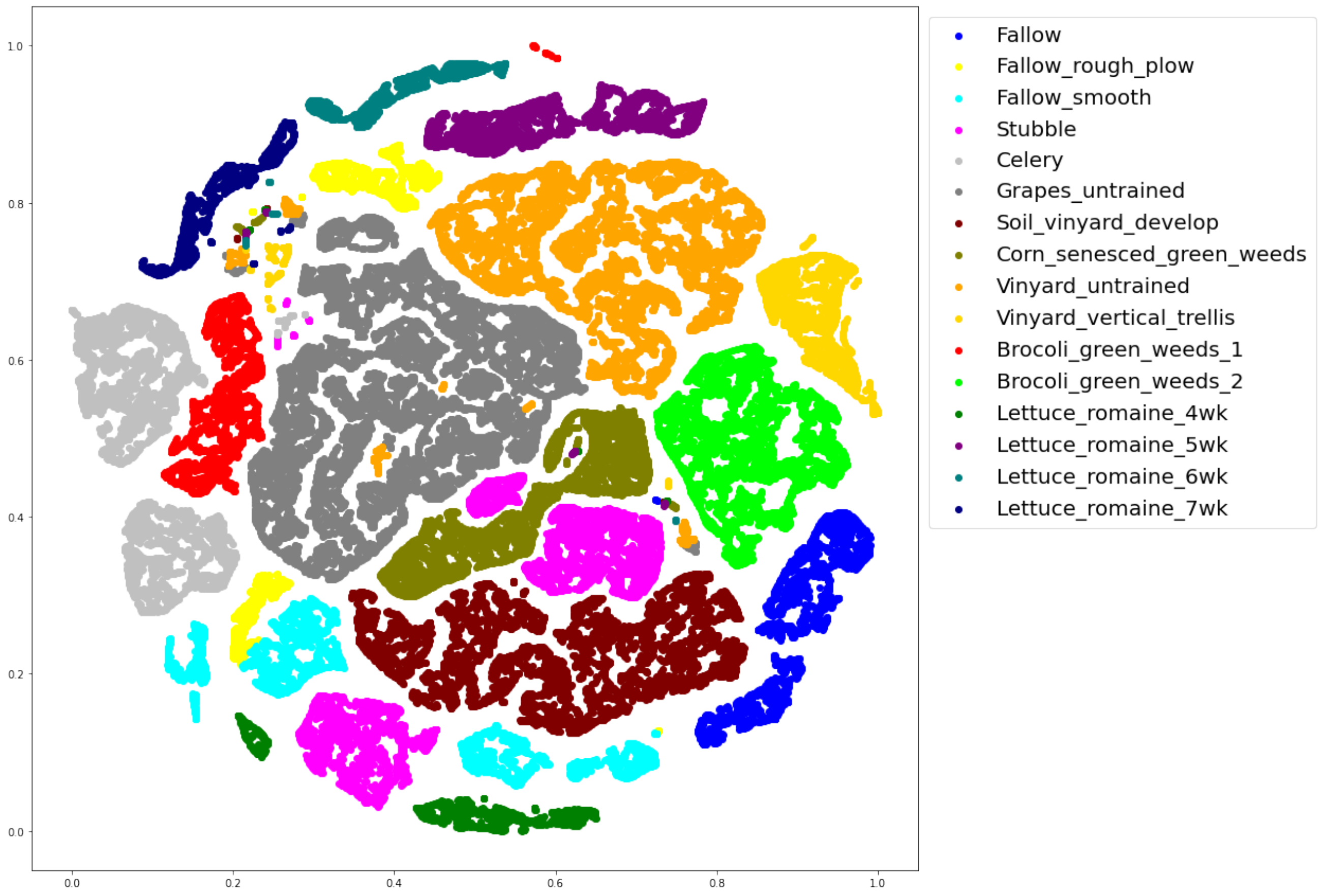}%
\label{fig:sa_features}}
\caption{Feature spaces learned for the known classes via self-supervised training (with VICReg \cite{vicreg}) for all 3 datasets. Although the classes are identified in the plot for contextualization, no annotation is used in the SSL training. (a) Indian Pines. (b) Kennedy Space Center. (c) Salinas.}
\label{fig:feature_space}
\end{figure}

Generating an adequate feature space can be the key to good performance in the prediction phase. This is especially true if the pixel prediction algorithm that follows the SSL training does not require further training, such as the proposed approaches, because the learned feature space is the foundation of the method in those cases. Figure \ref{fig:feature_space} shows the t-Distributed Stochastic Neighbor Embedding (t-SNE) \cite{t-sne} plot of the labeled data for the three evaluated datasets after the SSL training with VICReg \cite{vicreg}, where each point represents a labeled pixel of the input scene. The t-SNE representation was computed based on the output of the encoder backbone of the SSL method, whose dimension is of size 512, as detailed in Section \ref{sec:experimental}, instead of the output of the projection head that is used only for loss calculation during the SSL training.

\begin{figure}[ht!]
\centering
\subfloat[]{\includegraphics[width=0.75\linewidth]{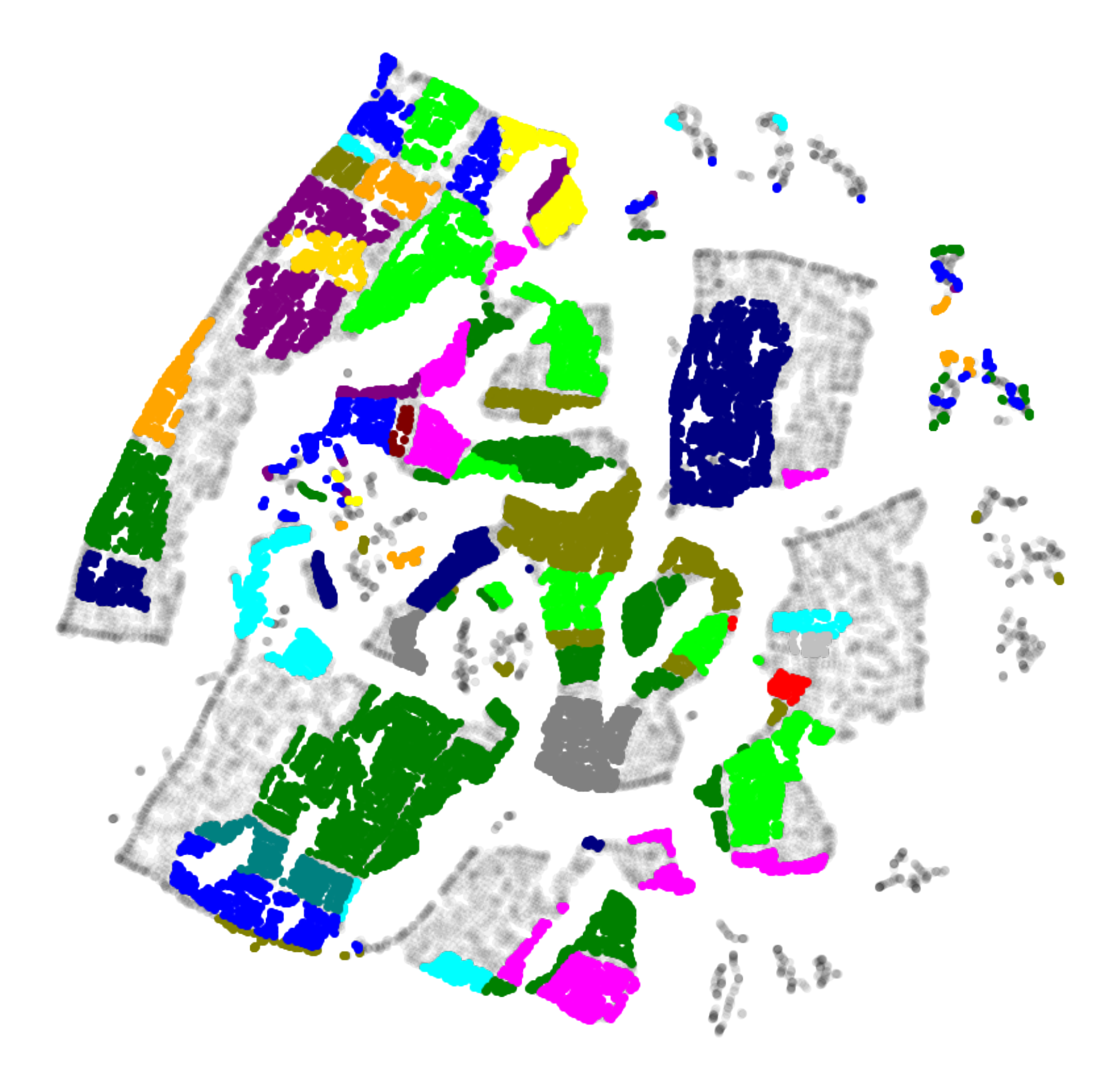}%
\label{fig:ip_features_all}}
\\
\subfloat[]{\includegraphics[width=0.75\linewidth]{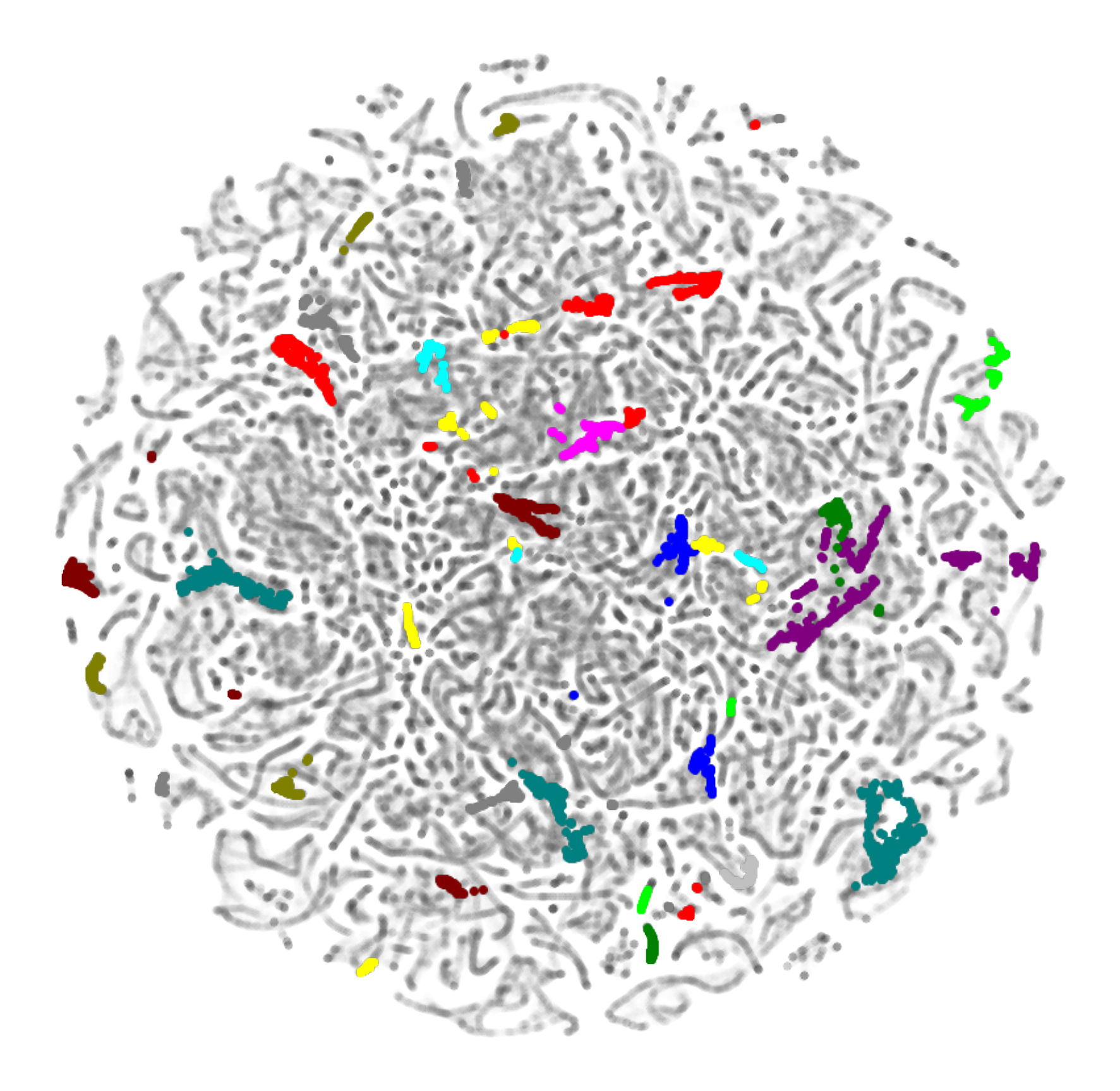}%
\label{fig:ksc_features_all}}
\\
\subfloat[]{\includegraphics[width=0.75\linewidth]{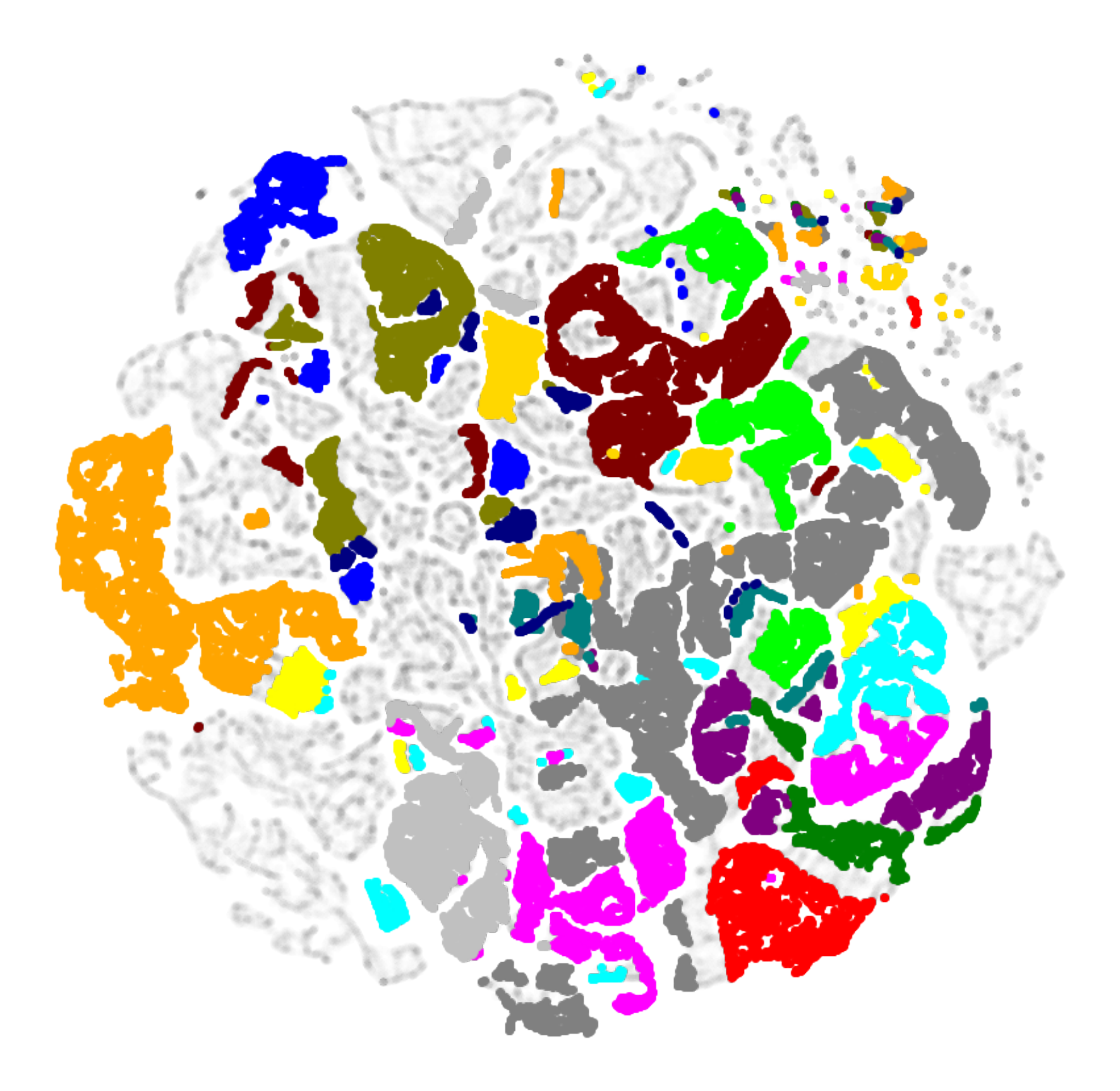}%
\label{fig:sa_features_all}}
\caption{Feature spaces learned using all pixels, labeled or not, via self-supervised training (with VICReg \cite{vicreg}) for all 3 datasets. Each light-gray point represents an unlabeled pixel. (a) Indian Pines. (b) Kennedy Space Center. (c) Salinas.}
\label{fig:feature_space_all}
\end{figure}

Figure \ref{fig:feature_space} shows an important outcome of the self-supervised training: the labeled classes tend to form separated clusters, a desirable characteristic to minimize misclassifications later in the pipeline. It is important to note that this class separation was achieved without any annotation. Therefore, the model organized this feature space using only the intrinsic spectral information of each pixel, combined with the local spatial context from its patch neighbors. The plots in Figure~\ref{fig:feature_space} focus on showing the overall organization of only the labeled pixels from the input HSI. However, in practice, the unlabeled pixels usually distribute themselves around the clusters of known classes, either for being other samples from known classes that were not labeled, or for being a sample of an unknown class that could not be grouped with other clusters. This property can be seen in Figure~\ref{fig:feature_space_all}, where the unlabeled pixels are represented by the light-gray points. This aligns with the physical reality of remote sensing, as these intermediate points likely represent boundary pixels or mixed pixels created by the sensor's limited spatial resolution. Consequently, while the labeled class clusters are usually distinct, these surrounding unlabeled points create a smooth transition between them. This illustrates the main challenge of the task: defining a robust decision boundary that can navigate this dense and continuous transition region.

As the three datasets present classes whose pixels are usually contained in continually annotated clusters, this results in clusters of classes better separated in the feature space. It is important to note that datasets such as IP and SA present classes with close spatial proximity and distant semantic proximity that were well separated in the feature space nonetheless. An example of this in the IP dataset can be seen with the woods (dark blue) being separated in the feature space (Figure \ref{fig:ip_features}) from the classes corn-no till (light green), soybeans-no till (olive) and soybeans-min till (dark green), which are spatially close (as viewed on the top right quadrant of Figure \ref{fig:ip_label}). For SA, this can be seen with the class corn senesced green weeds (olive) being separated in the feature space of Figure \ref{fig:sa_features} from the classes lettuce romaine 7wk (navy/darker blue) and soil vineyard develop (dark red), although being spatially close, as seen Figure \ref{fig:sa_label} (below the middle of the image). This indicates that FOSSeL can work well even in scenarios where one needs to classify all pixels of the input image.

\subsection{Comparison to Unsupervised Graph-Propagation-Based Baselines}

From the results obtained by the SSL training for the generation of the feature space, FOSSeL performs pixel prediction from a few labeled pixels for each known class. In Tables \ref{tab:quantitative_wc_5}, \ref{tab:quantitative_wc_10} and  \ref{tab:quantitative_wc_15}, we show the comparison between the results obtained by FOSSeL and the baseline Watershed Classifier \cite{challa2019watersheds}, with and without employing the proposed superpixel strategy along with the label propagation algorithm. $\textit{OA Open}$ represents the overall accuracy of the model considering the existence of unknown classes, \textit{OA Closed} represents the accuracy of the model evaluated only on the known classes, \textit{BA Open} is the balanced accuracy considering the presence of the unknown class, \textit{BA Closed} is the balanced accuracy over the known classes only, and \textit{Kappa} is the kappa score including the unknown class. Additionally, in open-set scenarios, these traditional metrics can be limited in the amount of information they bring about the results, particularly when there is a significant imbalance between the number of known and unknown pixels. A desirable approach for an open-set method is to maintain a good balance between correctly classifying the known pixels and detecting unknown samples. Thus, we also adopt the H-score \cite{h-score} as a main performance indicator. The H-score calculates the harmonic mean between the classification accuracy of known classes and the accuracy of unknown classes, similar to the classic F1-Score, but adapted for the context of open-set problems. The H-score is defined as follows \cite{h-score}:

\begin{equation}
    \text{H-score} = 2 \cdot \frac{ACC_{known} \cdot ACC_{unknown}}{ACC_{known} + ACC_{unknown}} 
\end{equation}

The harmonic mean heavily penalizes large disparities between the two values: known accuracy and unknown accuracy. Consequently, a high H-score can only be achieved if the model successfully maintains high discriminative power for known classes without sacrificing its ability to identify unknown pixels. This prevents scenarios where a model artificially inflates its accuracy by over-classifying pixels into known categories. Thus, we focus our analysis on the $H$-score, as it provides a balanced metric to evaluate the alterations in the results according to hyperparameter changes.

All the following results were computed, respectively, by using only 5, 10 and 15 labeled pixels for label propagation. For each case, we ran the experiments 10 times, thus, we also report the metrics in terms of their standard deviation.

As Tables \ref{tab:quantitative_wc_5}, \ref{tab:quantitative_wc_10} and \ref{tab:quantitative_wc_15} show, the Watershed Classifier (WC) and SGLP consistently achieve worse results than the two best options among the proposed methods, with the WC ranking at best as the second highest option in terms of H-Score in 3 of the 9 experiments. Considering WC, this is directly tied to its criteria when calculating the \textit{pass value}. Because the Watershed Classifier determines the \textit{pass value} based on the minimum of the highest weights between vertices along a path, it frequently encounters the same bottleneck edges for different classes. This limitation results in a high tendency to generate identical pass values for different competing classes, leading to ambiguity and reduced discriminative power during the final classification step. 

As for SGLP, its comparatively poor overall performance can be primarily attributed to its reliance on the original hyperspectral data space for label propagation. Because SGLP computes similarities directly on the original pixel data rather than on a pre-trained feature space, it can hardly deal with complex spectral-spatial relationships. This limitation strongly reinforces the importance of employing FOSSeL's self-supervised learning (SSL) pre-training. Despite lacking this pre-training step, SGLP shows an overall performance similar to the WC: on the IP dataset, its results align with the WC w/sp variant, and it also achieves overall superior results compared to both WC versions on the SA dataset. However, the absence of a refined, self-supervised feature space proves critically detrimental on the KSC dataset, where the algorithm completely fails to accurately identify any unknown pixel, yielding an H-score of 0 across all evaluated cases.

A comparative analysis between the standard Watershed Classifier (WC) and its superpixel-enhanced variant (WC w/sp) illustrates the advantage of performing some type of spatial regularization in HSI pixel-wise classification. As Tables \ref{tab:quantitative_wc_5}, \ref{tab:quantitative_wc_10} and \ref{tab:quantitative_wc_15} show, the standard WC achieves poor results across the majority of evaluated scenarios when considering the BA Open metric as the main indicator, particularly within the IP and SA datasets. However, the addition of the superpixel segmentation (WC w/sp) brings a performance improvement in these cases. For example, in the SA 15-shot scenario, the BA Open increases from 66.25\% with the standard WC to 88.20\% with WC w/sp. This difference shows that pixel-wise graph propagation can be susceptible to local noise and spectral variability inherent to hyperspectral data. Hence, the superpixel segmentation step acts as a regularizer, reducing the effect of local anomalies and establishing a more balanced (and smaller) graph topology. This observation validates our framework's reliance on superpixel segmentation to stabilize the underlying graph before propagation begins. However, this strategy can also worsen the performance, as seen in WC for the KSC dataset, especially in cases where neighboring classes get combined in a single superpixel, misclassifying part of the pixels from the group depending on the centroid's prediction. On the other hand, by considering the H-Score as the main metric, we can see that the standard pixel-wise approach consistently yields more balanced results between known and unknown accuracy for 10- and 15-shot scenarios. This is especially evident in the KSC dataset, where H-Score drops to zero in those cases, indicating that no unknown pixel was correctly isolated, whereas the known classes are being classified correctly, which can be observed with the higher BA Closed metric values. This reinforces the fact that superpixel segmentation can compromise the classification when neighboring pixels from different classes get combined in the same segment.

Unlike the Watershed Classifier, which can have its results heavily compromised in superpixel-based graphs due to pass-value bottlenecking, the proposed methods demonstrated robust performance across all datasets and few-shot configurations. Furthermore, comparing our approach to SGLP highlights the important role of the SSL pre-training for open-set prediction. Because SGLP relies on original raw data for propagation, it often fails to separate complex classes effectively from the unknown regions. In contrast, our method operates on a more refined, self-supervised feature space that improves prediction robustness. These results highlight the importance of combining rich feature representations with spatial regularization in hyperspectral pixel-wise classification. By grouping pixels into coherent regions, superpixels reduce the impact of local spectral variability and noise, enabling more stable label propagation. At the same time, this introduces a trade-off, as incorrect segmentation may propagate errors within a region.

Additionally, we can also assess the computational efficiency gained by the superpixel segmentation phase. For this, we compared the inference times of the proposed superpixel-based graph against a standard pixel-level graph. We evaluated the most computationally demanding scenario: the highest-resolution dataset (KSC) under the 15-shot configuration. Operating strictly at the pixel level generates a massive graph comprising 314,368 nodes (as it is an image of resolution $512 \times 614$) and 627,610 edges, which drastically inflates the parameter quantity during adjacency matrix construction. When instantiated using the NetworkX Python library \cite{networkx}, this pixel-wise graph consumes approximately 360 MB of memory. While this memory consumption is well within the limits of modern hardware, the main bottleneck of graph-based label propagation is the time complexity required to traverse all of the interconnected nodes. By contrast, the proposed superpixel-based graph (configured to generate approximately 5,000 SLIC \cite{slic} segments) compresses the topology to just 4,651 nodes and 10,284 edges, which represents a reduction of over 98\% in both graph dimensions, also reducing the memory usage to a mere 5.64 MB. This structural compression translates to important reductions in runtime: for the KNNProp method, the pixel-level graph requires approximately 2327 seconds (almost 40 minutes) to propagate labels, whereas the superpixel graph executes the same task in just 4 seconds. Similarly, our proposed CSProp method drops from 13 seconds on the pixel-level graph to a highly efficient 1.6 seconds on the superpixel graph. SimProp was the only exception, as it remained computationally bottlenecked at approximately 60 seconds in both configurations, indicating a heavy algorithmic overhead independent of the graph size. These metrics demonstrate that the superpixel strategy is a necessary methodological mechanism to make graph-based open-set pixel-wise classification memory-efficient and computationally viable for high-resolution hyperspectral imagery.

While CSProp establishes itself as the best algorithm across most datasets when focusing on the BA Open metric, by considering the H-Score as the main metric, SimProp and KNNProp consistently rank higher due to better maintaining the known/unknown accuracy balance. This is especially true for the IP dataset, where not only does CSProp rank below the other two FOSSeL variants, but it also shows a high standard deviation for the H-Score. The IP dataset is challenging due to its high concentration of mixed pixels and intra-class variance. In such environments, the explicit spatial-feature exploitation utilized by CSProp may propagate noise. On the other hand, SimProp relies on a smoother propagation, mitigating part of the local ambiguities present in the IP data.

However, the superiority of the CSProp method in terms of BA Open can be explained by the fact that CSProp explicitly leverages both the spatial structure of the graph and the feature-space distances to labeled seeds. By relying on shortest paths, the method effectively constrains label propagation to spatially consistent regions while still accounting for spectral similarity, resulting in more robust predictions in few-shot scenarios, especially for the known classes.

KNNProp was also shown to be a strong alternative, particularly on the SA dataset, where it achieves the best results in all of the scenarios. This is the pixel prediction algorithm that relies most heavily on feature proximity within the SSL space rather than explicit spatial information. Thus, its strong performance helps to demonstrate that the unsupervised features can be highly discriminative by themselves, which is particularly valuable for real-world scenarios where no label information may be available. However, the superiority of SimProp and CSProp over KNNProp in other scenarios confirms that working with spatial knowledge along with SSL features tends to yield the most accurate results.

\begin{table*}[ht]
\centering
\caption{Comparison between the results of the proposed FOSSeL method with varying pixel prediction algorithms, the Watershed Classifier (WC) \cite{challa2019watersheds}, WC with superpixels (WC w/sp), and SGLP \cite{sglp} in a 5-shot scenario, where only 5 labeled seeds are available for each known class. The best results in terms of H-Score for each dataset are in bold, and the second-best results are in red.}
\label{tab:quantitative_wc_5}
\resizebox{0.7\linewidth}{!}{%
{\def\arraystretch{1.2}
\begin{tabular}{clcccccc}
\hline
Dataset & \multicolumn{1}{c}{Method} & OA Open & OA Closed & BA Open & BA Closed & Kappa & H-Score \\ \hline
 & WC & $28.21 \pm 3.27$ & $27.85 \pm 3.34$ & $35.91 \pm 3.35$ & $33.48 \pm 3.47$ & $23.96 \pm 3.17$ & $38.44 \pm 4.07$ \\
 & WC w/sp &  $44.64 \pm 3.96$ & $44.56 \pm 4.06$ & $56.60 \pm 2.70$ & $56.80 \pm 2.34$ & $39.85 \pm 3.99$ & $46.52 \pm 7.76$ \\
 & SGLP & $51.44 \pm 4.99$ & $51.49 \pm 4.97$ & $60.02 \pm 4.49$ & $61.13 \pm 4.10$ & $46.45 \pm 5.14$ & $45.07 \pm 16.63$ \\
 & CSProp (ours) &  $63.63 \pm 2.05$ & $63.87 \pm 2.00$ & $72.88 \pm 1.97$ & $75.57 \pm 1.58$ & $60.21 \pm 2.10$ & $43.29 \pm 22.22$ \\
 & \textbf{SimProp (ours)} &  $43.86 \pm 2.90$ & $43.36 \pm 2.92$ & $62.19 \pm 2.13$ & $59.48 \pm 2.35$ & $40.99 \pm 2.76$ & $59.86 \pm 3.38$ \\
\multirow{-6}{*}{IP} & \textcolor{red}{KNNProp (ours)} &  $41.31 \pm 3.62$ & $40.85 \pm 3.65$ & $58.40 \pm 2.83$ & $55.92 \pm 3.06$ & $38.39 \pm 3.48$ & $55.96 \pm 4.55$  \\ \hline
 & WC &  $76.94 \pm 2.21$ & $73.14 \pm 2.26$ & $67.81 \pm 2.62$ & $65.51 \pm 3.24$ & $73.27 \pm 2.42$ & $80.78 \pm 3.53$ \\
 & WC w/sp &  $68.34 \pm 1.78$ & $86.37 \pm 2.49$ & $74.88 \pm 1.87$ & $82.14 \pm 2.16$ & $65.15 \pm 1.97$ & $4.50 \pm 2.94$ \\
 & SGLP & $56.92 \pm 5.86$ & $72.31 \pm 7.45$ & $62.11 \pm 4.33$ & $68.32 \pm 4.77$ & $52.62 \pm 6.51$ & $0.00 \pm 0.00$ \\
 & \textcolor{red}{CSProp (ours)} & $90.31 \pm 1.52$ & $97.43 \pm 1.97$ & $93.61 \pm 1.38$ & $96.55 \pm 1.55$ & $89.10 \pm 1.70$ & $77.38 \pm 1.45$   \\
 & \textbf{SimProp (ours)} & $88.09 \pm 2.77$ & $85.61 \pm 3.36$ & $84.06 \pm 1.54$ & $82.75 \pm 1.65$ & $86.19 \pm 3.22$ & $90.99 \pm 2.19$  \\
\multirow{-6}{*}{KSC} & KNNProp (ours) &   $79.67 \pm 1.36$ & $81.35 \pm 1.62$ & $74.18 \pm 1.06$ & $74.25 \pm 1.33$ & $76.52 \pm 1.55$ & $77.02 \pm 3.22$ \\ \hline
 & WC &  $66.06 \pm 2.85$ & $67.62 \pm 1.48$ & $62.31 \pm 3.58$ & $63.28 \pm 2.49$ & $62.18 \pm 3.23$ & $54.60 \pm 15.82$ \\
 & \textcolor{red}{WC w/sp} &  $84.01 \pm 1.97$ & $84.34 \pm 2.19$ & $86.50 \pm 1.70$ & $86.98 \pm 1.85$ & $82.18 \pm 2.20$ & $82.49 \pm 1.00$ \\
 & SGLP & $67.64 \pm 5.11$ & $70.21 \pm 5.53$ & $72.11 \pm 2.88$ & $74.59 \pm 3.12$ & $63.94 \pm 5.58$ & $52.70 \pm 3.32$ \\
 & CSProp (ours) & $85.47 \pm 2.19$ & $87.13 \pm 2.24$ & $87.35 \pm 1.45$ & $88.88 \pm 1.47$ & $83.84 \pm 2.40$ & $76.93 \pm 3.81$ \\
 & SimProp (ours) & $70.08 \pm 2.38$ & $69.51 \pm 2.52$ & $72.60 \pm 1.48$ & $72.34 \pm 1.46$ & $66.82 \pm 2.56$ & $72.20 \pm 4.08$  \\
\multirow{-6}{*}{SA} & \textbf{KNNProp (ours)} &  $86.58 \pm 4.06$ & $86.53 \pm 4.33$ & $88.34 \pm 1.59$ & $88.44 \pm 1.66$ & $85.03 \pm 4.43$ & $86.69 \pm 3.72$ \\ \hline
\end{tabular}%
}
}
\end{table*}

\begin{table*}[ht]
\centering
\caption{Comparison between the results of the proposed FOSSeL method with varying pixel prediction algorithms, the Watershed Classifier (WC) \cite{challa2019watersheds}, WC with superpixels (WC w/sp), and SGLP \cite{sglp} in a 10-shot scenario, where only 10 labeled seeds are available for each known class. The best results in terms of H-Score for each dataset are in bold, and the second-best results are in red.}
\label{tab:quantitative_wc_10}
\resizebox{0.7\linewidth}{!}{%
{\def\arraystretch{1.2}
\begin{tabular}{clcccccc}
\hline
Dataset & \multicolumn{1}{c}{Method} & OA Open & OA Closed & BA Open & BA Closed & Kappa & H-Score \\ \hline
 & WC & $37.00 \pm 2.29$ & $36.78 \pm 2.42$ & $43.80 \pm 1.47$ & $42.46 \pm 1.86$ & $32.67 \pm 2.06$ & $45.03 \pm 2.63$  \\
 & WC w/sp & $56.61 \pm 3.01$ & $56.68 \pm 3.00$ & $68.26 \pm 2.10$ & $69.67 \pm 1.93$ & $52.46 \pm 3.17$ & $49.97 \pm 13.35$  \\
 & SGLP & $61.46 \pm 3.42$ & $61.66 \pm 3.44$ & $68.46 \pm 3.34$ & $70.68 \pm 3.07$ & $57.26 \pm 3.61$ & $43.27 \pm 21.83$ \\
 & CSProp (ours) & $74.19 \pm 2.67$ & $74.59 \pm 2.69$ & $80.18 \pm 1.94$ & $83.94 \pm 1.78$ & $71.44 \pm 2.81$ & $41.36 \pm 17.50$ \\
  & \textbf{SimProp (ours)} & $64.21 \pm 3.17$ & $63.94 \pm 3.18$ & $76.71 \pm 2.17$ & $75.48 \pm 2.15$ & $61.33 \pm 3.18$ & $75.36 \pm 5.06$ \\
\multirow{-6}{*}{IP} & \textcolor{red}{KNNProp (ours)} &  $60.96 \pm 3.21$ & $60.72 \pm 3.22$ & $74.19 \pm 2.04$ & $73.27 \pm 1.89$ & $57.96 \pm 3.19$ & $70.41 \pm 7.39$ \\ \hline
 & \textcolor{red}{WC} & $82.33 \pm 1.49$ & $78.49 \pm 1.48$ & $74.03 \pm 1.52$ & $71.82 \pm 1.65$ & $79.43 \pm 1.70$ & $86.40 \pm 1.81$ \\
 & WC w/sp & $71.21 \pm 0.00$ & $90.71 \pm 0.00$ & $78.27 \pm 0.00$ & $86.10 \pm 0.00$ & $68.42 \pm 0.00$ & $0.00 \pm 0.00$ \\
 & SGLP & $63.73 \pm 3.49$ & $80.97 \pm 4.43$ & $70.93 \pm 2.66$ & $78.02 \pm 2.92$ & $60.13 \pm 3.87$ & $0.00 \pm 0.00$ \\
 & CSProp (ours) & $91.30 \pm 0.03$ & $98.51 \pm 0.05$ & $94.44 \pm 0.04$ & $97.38 \pm 0.05$ & $90.21 \pm 0.03$ & $78.32 \pm 0.02$  \\
 & \textbf{SimProp (ours)} &  $90.56 \pm 1.27$ & $89.51 \pm 1.62$ & $85.95 \pm 0.65$ & $85.11 \pm 0.71$ & $89.10 \pm 1.47$ & $91.89 \pm 0.87$  \\
\multirow{-6}{*}{KSC} & KNNProp (ours) & $80.42 \pm 0.00$ & $83.21 \pm 0.00$ & $75.42 \pm 0.00$ & $75.94 \pm 0.00$ & $77.45 \pm 0.00$ & $76.18 \pm 0.00$ \\ \hline
 & WC & $67.43 \pm 1.58$ & $70.25 \pm 1.20$ & $64.95 \pm 1.86$ & $67.06 \pm 1.49$ & $63.77 \pm 1.80$ & $49.62 \pm 8.59$ \\
 & WC w/sp & $85.12 \pm 4.29$ & $89.66 \pm 1.44$ & $88.38 \pm 3.32$ & $92.37 \pm 0.83$ & $83.44 \pm 4.76$ & $43.08 \pm 41.83$  \\
 & SGLP & $71.45 \pm 4.10$ & $74.41 \pm 4.52$ & $75.09 \pm 2.09$ & $77.83 \pm 2.31$ & $68.16 \pm 4.45$ & $53.81 \pm 2.27$ \\
 & \textcolor{red}{CSProp (ours)} &  $91.86 \pm 1.59$ & $93.58 \pm 1.43$ & $93.95 \pm 1.09$ & $95.53 \pm 1.04$ & $90.93 \pm 1.76$ & $83.15 \pm 3.81$ \\
 & SimProp (ours) & $83.96 \pm 1.59$ & $87.00 \pm 1.63$ & $85.73 \pm 1.35$ & $88.37 \pm 1.38$ & $82.12 \pm 1.75$ & $66.62 \pm 2.93$ \\
\multirow{-6}{*}{SA} & \textbf{KNNProp (ours)} & $91.44 \pm 1.87$ & $92.67 \pm 2.06$ & $93.05 \pm 0.88$ & $94.19 \pm 0.97$ & $90.44 \pm 2.06$ & $85.44 \pm 3.06$  \\ \hline
\end{tabular}
}
}
\end{table*}

\begin{table*}[ht]
\centering
\caption{Comparison between the results of the proposed FOSSeL method with varying pixel prediction algorithms, the Watershed Classifier (WC) \cite{challa2019watersheds}, WC with superpixels (WC w/sp), and SGLP \cite{sglp} in a 15-shot scenario, where only 15 labeled seeds are available for each known class. The best results in terms of H-Score for each dataset are in bold, and the second-best results are in red.}
\label{tab:quantitative_wc_15}
\resizebox{0.7\linewidth}{!}{%
{\def\arraystretch{1.2}
\begin{tabular}{clcccccc}
\hline
Dataset & \multicolumn{1}{c}{Method} & OA Open & OA Closed & BA Open & BA Closed & Kappa & H-Score \\ \hline
 & WC & $42.65 \pm 3.21$ & $42.42 \pm 3.33$ & $49.80 \pm 1.57$ & $48.45 \pm 2.62$ & $38.34 \pm 3.20$ & $51.11 \pm 3.03$  \\
 & WC w/sp & $64.81 \pm 3.38$ & $65.10 \pm 3.35$ & $74.46 \pm 2.15$ & $77.57 \pm 1.59$ & $61.26 \pm 3.62$ & $43.43 \pm 10.78$  \\
 & SGLP & $67.24 \pm 2.38$ & $67.56 \pm 2.33$ & $73.92 \pm 2.83$ & $77.11 \pm 1.98$ & $63.62 \pm 2.58$ & $38.81 \pm 22.35$ \\
 & CSProp (ours) & $79.32 \pm 1.87$ & $79.68 \pm 1.76$ & $84.52 \pm 1.93$ & $87.87 \pm 0.82$ & $77.05 \pm 2.01$ & $52.19 \pm 15.44$ \\
 & \textbf{SimProp (ours)} &  $65.68 \pm 2.10$ & $65.41 \pm 2.13$ & $78.41 \pm 0.80$ & $77.16 \pm 0.91$ & $62.90 \pm 2.08$ & $77.28 \pm 2.16$ \\
\multirow{-6}{*}{IP} & \textcolor{red}{KNNProp (ours)} & $64.45 \pm 2.09$ & $64.14 \pm 2.11$ & $77.53 \pm 1.24$ & $76.04 \pm 1.27$ & $61.59 \pm 2.11$ & $77.09 \pm 2.21$  \\ \hline
 & \textcolor{red}{WC} &  $85.20 \pm 1.03$ & $82.04 \pm 0.95$ & $77.97 \pm 1.22$ & $76.12 \pm 1.27$ & $82.79 \pm 1.18$ & $88.67 \pm 1.25$ \\
 & WC w/sp & $71.18 \pm 0.00$ & $90.70 \pm 0.00$ & $78.27 \pm 0.00$ & $86.10 \pm 0.00$ & $68.39 \pm 0.00$ & $0.00 \pm 0.00$ \\
 & SGLP & $66.80 \pm 2.38$ & $84.86 \pm 3.02$ & $75.16 \pm 2.03$ & $82.68 \pm 2.23$ & $63.52 \pm 2.65$ & $0.00 \pm 0.00$ \\
 & CSProp (ours) & $91.29 \pm 0.00$ & $98.49 \pm 0.00$ & $94.42 \pm 0.00$ & $97.36 \pm 0.00$ & $90.19 \pm 0.00$ & $78.33 \pm 0.00$ \\
 & \textbf{SimProp (ours)} & $88.73 \pm 0.00$ & $86.30 \pm 0.00$ & $82.39 \pm 0.00$ & $80.87 \pm 0.00$ & $86.90 \pm 0.00$ & $91.59 \pm 0.00$  \\
\multirow{-6}{*}{KSC} & KNNProp (ours) &  $80.42 \pm 0.00$ & $82.76 \pm 0.00$ & $74.89 \pm 0.00$ & $75.19 \pm 0.00$ & $77.38 \pm 0.00$ & $76.93 \pm 0.00$ \\ \hline
 & WC & $68.12 \pm 1.09$ & $71.93 \pm 0.70$ & $66.25 \pm 1.13$ & $69.22 \pm 0.91$ & $64.55 \pm 1.17$ & $41.86 \pm 11.65$ \\
 & WC w/sp & $84.70 \pm 3.11$ & $91.65 \pm 1.15$ & $88.20 \pm 2.46$ & $94.18 \pm 0.81$ & $82.96 \pm 3.46$ & $17.90 \pm 33.73$  \\
 & SGLP & $74.26 \pm 3.63$ & $77.47 \pm 3.99$ & $77.42 \pm 2.32$ & $80.32 \pm 2.56$ & $71.26 \pm 4.00$ & $54.88 \pm 2.19$ \\
 & \textcolor{red}{CSProp (ours)} & $94.43 \pm 0.60$ & $96.39 \pm 0.57$ & $96.04 \pm 0.52$ & $97.78 \pm 0.46$ & $93.78 \pm 0.67$ & $84.41 \pm 1.74$   \\
 & SimProp (ours) &  $86.03 \pm 1.36$ & $87.41 \pm 1.47$ & $88.73 \pm 1.06$ & $90.09 \pm 1.17$ & $84.42 \pm 1.50$ & $79.16 \pm 1.80$  \\
\multirow{-6}{*}{SA} & \textbf{KNNProp (ours)} & $92.98 \pm 1.33$ & $92.70 \pm 1.40$ & $94.66 \pm 0.77$ & $94.57 \pm 0.82$ & $92.16 \pm 1.47$ & $94.18 \pm 1.15$ \\ \hline
\end{tabular}%
}
}
\end{table*}

\subsection{Comparison to Supervised Baselines}

In Tables \ref{tab:quantitative_5}, \ref{tab:quantitative_10} and  \ref{tab:quantitative_15}, we show the comparison between the results obtained by FOSSeL and the supervised baselines OSC-SCL \cite{osc-scl}, SSMLP-RPL \cite{ssmlp-rpl}, HyperCASR \cite{hypercasr} and CACL \cite{cacl}. The experiments follow a similar pattern to the comparisons with the Watershed Classifier, thus, the results were computed, respectively, by using only 5, 10 and 15 labeled pixels for training (in the case of the baselines) or label propagation (in the case of FOSSeL) over 10 repetitions.

\begin{table*}[ht]
\centering
\caption{Comparison between the results of the proposed FOSSeL method with varying pixel prediction algorithms and the supervised baselines in a 5-shot scenario, where only 5 labeled pixels are available for each known class. The best results in terms of H-Score for each dataset are in bold, and the second-best results are in red.}
\label{tab:quantitative_5}
\resizebox{0.7\linewidth}{!}{%
{\def\arraystretch{1.2}
\begin{tabular}{clcccccc}
\hline
Dataset & \multicolumn{1}{c}{Method} & OA Open & OA Closed & BA Open & BA Closed & Kappa & H-Ccore\\ \hline
 & OSC-SCL \cite{osc-scl} & $41.29 \pm 2.00$ & $40.96 \pm 2.01$ & $55.91 \pm 1.95$ & $54.32 \pm 1.90$ & $37.70 \pm 1.86$ & $53.26 \pm 2.72$ \\
 & SSMLP-RPL \cite{ssmlp-rpl} & $41.49 \pm 4.85$ & $41.25 \pm 4.85$ & $53.08 \pm 4.54$ & $51.57 \pm 4.35$ & $38.18 \pm 4.53$ & $52.43 \pm 5.65$ \\
 & HyperCASR \cite{hypercasr} & $56.13 \pm 5.20$ & $56.65 \pm 5.24$ & $61.60 \pm 3.27$ & $66.30 \pm 3.49$ & $50.72 \pm 5.40$ & $1.01 \pm 2.41$ \\
 & \textbf{CACL \cite{cacl}} & $69.73 \pm 3.72$ & $69.79 \pm 3.78$ & $75.85 \pm 2.49$ & $76.78 \pm 2.32$ & $66.71 \pm 3.79$ & $65.01 \pm 10.61$ \\
 & CSProp (ours) & $63.63 \pm 2.05$ & $63.87 \pm 2.00$ & $72.88 \pm 1.97$ & $75.57 \pm 1.58$ & $60.21 \pm 2.10$ & $43.29 \pm 22.22$ \\
 & \textcolor{red}{SimProp (ours)} &  $43.86 \pm 2.90$ & $43.36 \pm 2.92$ & $62.19 \pm 2.13$ & $59.48 \pm 2.35$ & $40.99 \pm 2.76$ & $59.86 \pm 3.38$ \\
\multirow{-8}{*}{IP} & KNNProp (ours) &  $41.31 \pm 3.62$ & $40.85 \pm 3.65$ & $58.40 \pm 2.83$ & $55.92 \pm 3.06$ & $38.39 \pm 3.48$ & $55.96 \pm 4.55$ \\ \hline
 & OSC-SCL \cite{osc-scl} & $35.99 \pm 9.85$ & $18.51 \pm 12.51$ & $20.52 \pm 6.84$ & $12.59 \pm 7.50$ & $19.59 \pm 12.97$ & $29.38 \pm 17.63$ \\
 & SSMLP-RPL \cite{ssmlp-rpl} & $67.41 \pm 5.43$ & $77.31 \pm 6.96$ & $68.05 \pm 6.40$ & $71.78 \pm 7.35$ & $63.21 \pm 6.22$ & $40.50 \pm 18.07$ \\
 & HyperCASR \cite{hypercasr} & $71.04 \pm 3.19$ & $88.64 \pm 2.49$ & $79.39 \pm 2.83$ & $86.66 \pm 2.95$ & $68.12 \pm 3.40$ & $10.75 \pm 15.34$ \\
 & CACL \cite{cacl} & $81.80 \pm 6.57$ & $91.37 \pm 2.68$ & $85.84 \pm 2.42$ & $89.74 \pm 3.16$ & $79.61 \pm 7.15$ & $55.34 \pm 28.09$ \\
 & \textcolor{red}{CSProp (ours)} & $90.31 \pm 1.52$ & $97.43 \pm 1.97$ & $93.61 \pm 1.38$ & $96.55 \pm 1.55$ & $89.10 \pm 1.70$ & $77.38 \pm 1.45$ \\
 & \textbf{SimProp (ours)} & $88.09 \pm 2.77$ & $85.61 \pm 3.36$ & $84.06 \pm 1.54$ & $82.75 \pm 1.65$ & $86.19 \pm 3.22$ & $90.99 \pm 2.19$ \\
\multirow{-8}{*}{KSC} & KNNProp (ours) &  $79.67 \pm 1.36$ & $81.35 \pm 1.62$ & $74.18 \pm 1.06$ & $74.25 \pm 1.33$ & $76.52 \pm 1.55$ & $77.02 \pm 3.22$ \\ \hline
 & OSC-SCL \cite{osc-scl} & $60.07 \pm 2.65$ & $60.59 \pm 3.19$ & $76.38 \pm 1.71$ & $78.17 \pm 1.84$ & $56.66 \pm 2.80$ & $56.49 \pm 7.15$ \\
 & SSMLP-RPL \cite{ssmlp-rpl} & $74.92 \pm 5.11$ & $75.68 \pm 5.85$ & $81.62 \pm 5.25$ & $82.80 \pm 6.06$ & $72.16 \pm 5.57$ & $69.90 \pm 10.39$  \\
 & HyperCASR \cite{hypercasr} & $73.58 \pm 1.70$ & $80.87 \pm 1.90$ & $80.52 \pm 2.01$ & $87.08 \pm 2.16$ & $70.63 \pm 1.84$ & $3.26 \pm 2.74$ \\
 & CACL \cite{cacl} & $80.79 \pm 5.06$ & $83.94 \pm 4.95$ & $85.33 \pm 3.91$ & $88.30 \pm 3.89$ & $78.65 \pm 5.56$ & $61.05 \pm 13.49$ \\
 & \textcolor{red}{CSProp (ours)} &  $85.47 \pm 2.19$ & $87.13 \pm 2.24$ & $87.35 \pm 1.45$ & $88.88 \pm 1.47$ & $83.84 \pm 2.40$ & $76.93 \pm 3.81$ \\
 & SimProp (ours) & $70.08 \pm 2.38$ & $69.51 \pm 2.52$ & $72.60 \pm 1.48$ & $72.34 \pm 1.46$ & $66.82 \pm 2.56$ & $72.20 \pm 4.08$ \\
\multirow{-8}{*}{SA} & \textbf{KNNProp (ours)} & $86.58 \pm 4.06$ & $86.53 \pm 4.33$ & $88.34 \pm 1.59$ & $88.44 \pm 1.66$ & $85.03 \pm 4.43$ & $86.69 \pm 3.72$  \\ \hline
\end{tabular}%
}
}
\end{table*}

\begin{table*}[ht]
\centering
\caption{Comparison between the results of the proposed FOSSeL method with varying pixel prediction algorithms and the supervised baselines in a 10-shot scenario, where only 10 labeled pixels are available for each known class. The best results in terms of H-Score for each dataset are in bold, and the second-best results are in red.}
\label{tab:quantitative_10}
\resizebox{0.7\linewidth}{!}{%
{\def\arraystretch{1.2}
\begin{tabular}{clcccccc}
\hline
Dataset & \multicolumn{1}{c}{Method} & OA Open & OA Closed & BA Open & BA Closed & Kappa & H-Score\\ \hline
 & OSC-SCL \cite{osc-scl} & $56.05 \pm 1.60$ & $55.87 \pm 1.60$ & $71.74 \pm 1.01$ & $71.50 \pm 0.92$ & $53.03 \pm 1.55$ & $63.91 \pm 2.51$ \\
 & SSMLP-RPL \cite{ssmlp-rpl} & $55.69 \pm 4.06$ & $55.58 \pm 4.08$ & $65.53 \pm 3.39$ & $64.96 \pm 3.72$ & $52.54 \pm 3.94$ & $62.77 \pm 4.85$ \\
 & HyperCASR \cite{hypercasr} & $66.54 \pm 3.25$ & $67.14 \pm 3.29$ & $71.20 \pm 1.73$ & $76.44 \pm 1.99$ & $62.37 \pm 3.54$ & $5.57 \pm 6.16$ \\
 & \textcolor{red}{CACL \cite{cacl}} & $85.59 \pm 2.21$ & $85.79 \pm 2.18$ & $87.12 \pm 2.34$ & $88.93 \pm 1.77$ & $83.80 \pm 2.43$ & $71.51 \pm 13.06$ \\
 & CSProp (ours) &  $74.19 \pm 2.67$ & $74.59 \pm 2.69$ & $80.18 \pm 1.94$ & $83.94 \pm 1.78$ & $71.44 \pm 2.81$ & $41.36 \pm 17.50$ \\
 & \textbf{SimProp (ours)} & $64.21 \pm 3.17$ & $63.94 \pm 3.18$ & $76.71 \pm 2.17$ & $75.48 \pm 2.15$ & $61.33 \pm 3.18$ & $75.36 \pm 5.06$ \\
\multirow{-8}{*}{IP} & KNNProp (ours) &  $60.96 \pm 3.21$ & $60.72 \pm 3.22$ & $74.19 \pm 2.04$ & $73.27 \pm 1.89$ & $57.96 \pm 3.19$ & $70.41 \pm 7.39$ \\ \hline
 & OSC-SCL \cite{osc-scl} & $62.78 \pm 5.48$ & $57.14 \pm 7.10$ & $48.60 \pm 6.75$ & $45.14 \pm 7.82$ & $55.29 \pm 6.85$ & $66.87 \pm 5.77$ \\
 & SSMLP-RPL \cite{ssmlp-rpl} & $74.85 \pm 3.07$ & $90.12 \pm 3.70$ & $80.77 \pm 3.55$ & $87.03 \pm 3.95$ & $72.03 \pm 3.45$ & $29.77 \pm 8.56$ \\
 & HyperCASR \cite{hypercasr} & $74.82 \pm 1.23$ & $94.60 \pm 1.25$ & $85.61 \pm 1.18$ & $93.83 \pm 1.26$ & $72.31 \pm 1.34$ & $6.42 \pm 6.12$ \\
 & CACL \cite{cacl} & $82.28 \pm 4.69$ & $97.49 \pm 1.57$ & $90.09 \pm 1.97$ & $96.36 \pm 2.64$ & $80.35 \pm 5.07$ & $37.40 \pm 26.16$ \\
 & \textcolor{red}{CSProp (ours)} &  $91.30 \pm 0.03$ & $98.51 \pm 0.05$ & $94.44 \pm 0.04$ & $97.38 \pm 0.05$ & $90.21 \pm 0.03$ & $78.32 \pm 0.02$ \\
 & \textbf{SimProp (ours)} & $90.56 \pm 1.27$ & $89.51 \pm 1.62$ & $85.95 \pm 0.65$ & $85.11 \pm 0.71$ & $89.10 \pm 1.47$ & $91.89 \pm 0.87$ \\
\multirow{-8}{*}{KSC} & KNNProp (ours) & $80.42 \pm 0.00$ & $83.21 \pm 0.00$ & $75.42 \pm 0.00$ & $75.94 \pm 0.00$ & $77.45 \pm 0.00$ & $76.18 \pm 0.00$ \\ \hline
 & OSC-SCL \cite{osc-scl} &  $80.13 \pm 2.03$ & $84.73 \pm 0.83$ & $88.18 \pm 1.51$ & $92.63 \pm 0.57$ & $78.03 \pm 2.22$ & $45.50 \pm 21.61$ \\
 & SSMLP-RPL \cite{ssmlp-rpl} &  $81.50 \pm 3.11$ & $82.79 \pm 3.84$ & $86.94 \pm 1.83$ & $88.46 \pm 2.04$ & $79.39 \pm 3.39$ & $72.32 \pm 17.48$ \\
 & HyperCASR \cite{hypercasr} & $76.02 \pm 1.73$ & $83.32 \pm 1.93$ & $83.67 \pm 1.94$ & $90.30 \pm 2.09$ & $73.36 \pm 1.87$ & $7.66 \pm 5.95$  \\
 & CACL \cite{cacl} & $88.26 \pm 2.23$ & $92.78 \pm 2.14$ & $91.08 \pm 1.45$ & $95.02 \pm 1.07$ & $86.91 \pm 2.47$ & $57.48 \pm 16.13$ \\
 & \textcolor{red}{CSProp (ours)} &  $91.86 \pm 1.59$ & $93.58 \pm 1.43$ & $93.95 \pm 1.09$ & $95.53 \pm 1.04$ & $90.93 \pm 1.76$ & $83.15 \pm 3.81$ \\
 & SimProp (ours) & $83.96 \pm 1.59$ & $87.00 \pm 1.63$ & $85.73 \pm 1.35$ & $88.37 \pm 1.38$ & $82.12 \pm 1.75$ & $66.62 \pm 2.93$  \\
\multirow{-8}{*}{SA} & \textbf{KNNProp (ours)} & $91.44 \pm 1.87$ & $92.67 \pm 2.06$ & $93.05 \pm 0.88$ & $94.19 \pm 0.97$ & $90.44 \pm 2.06$ & $85.44 \pm 3.06$  \\ \hline
\end{tabular}
}
}
\end{table*}

\begin{table*}[ht]
\centering
\caption{Comparison between the results of the proposed FOSSeL method with varying pixel prediction algorithms and the supervised baselines in a 15-shot scenario, where only 15 labeled pixels are available for each known class. The best results in terms of H-Score for each dataset are in bold, and the second-best results are in red.}
\label{tab:quantitative_15}
\resizebox{0.7\linewidth}{!}{%
{\def\arraystretch{1.2}
\begin{tabular}{clcccccc}
\hline
Dataset & \multicolumn{1}{c}{Method} & OA Open & OA Closed & BA Open & BA Closed & Kappa & H-Score \\ \hline
 & OSC-SCL \cite{osc-scl} & $72.42 \pm 2.71$ & $72.50 \pm 2.72$ & $81.02 \pm 1.49$ & $82.31 \pm 1.35$ & $69.84 \pm 2.81$ & $67.88 \pm 5.06$ \\
 & SSMLP-RPL \cite{ssmlp-rpl} & $63.33 \pm 4.56$ & $63.29 \pm 4.60$ & $71.64 \pm 3.12$ & $71.71 \pm 3.60$ & $60.30 \pm 4.66$ & $66.42 \pm 4.40$ \\
 & HyperCASR \cite{hypercasr} & $73.45 \pm 2.04$ & $74.07 \pm 2.03$ & $76.89 \pm 1.50$ & $82.19 \pm 1.18$ & $70.14 \pm 2.17$ & $13.19 \pm 12.09$     \\
 & CACL \cite{cacl} & $90.81 \pm 2.04$ & $91.20 \pm 2.10$ & $89.87 \pm 1.43$ & $92.94 \pm 1.52$ & $89.61 \pm 2.27$ & $63.76 \pm 8.76$ \\
 & CSProp (ours) & $79.32 \pm 1.87$ & $79.68 \pm 1.76$ & $84.52 \pm 1.93$ & $87.87 \pm 0.82$ & $77.05 \pm 2.01$ & $52.19 \pm 15.44$ \\
 & \textbf{SimProp (ours)} & $65.68 \pm 2.10$ & $65.41 \pm 2.13$ & $78.41 \pm 0.80$ & $77.16 \pm 0.91$ & $62.90 \pm 2.08$ & $77.28 \pm 2.16$ \\
\multirow{-8}{*}{IP} & \textcolor{red}{KNNProp (ours)} & $64.45 \pm 2.09$ & $64.14 \pm 2.11$ & $77.53 \pm 1.24$ & $76.04 \pm 1.27$ & $61.59 \pm 2.11$ & $77.09 \pm 2.21$ \\ \hline
 & OSC-SCL \cite{osc-scl} & $65.00 \pm 4.33$ & $64.18 \pm 9.30$ & $53.71 \pm 7.34$ & $52.29 \pm 9.97$ & $58.54 \pm 5.28$ & $60.33 \pm 13.05$ \\
 & SSMLP-RPL \cite{ssmlp-rpl} & $79.52 \pm 4.61$ & $92.81 \pm 1.43$ & $84.75 \pm 2.05$ & $90.21 \pm 1.94$ & $77.16 \pm 5.00$ & $41.59 \pm 22.43$ \\
 & HyperCASR \cite{hypercasr} & $77.21 \pm 0.72$ & $97.28 \pm 1.03$ & $88.34 \pm 0.81$ & $96.61 \pm 0.99$ & $74.91 \pm 0.78$ & $10.63 \pm 4.45$ \\
 & CACL \cite{cacl} & $85.50 \pm 5.94$ & $99.12 \pm 0.73$ & $93.16 \pm 2.85$ & $98.78 \pm 0.93$ & $83.90 \pm 6.48$ & $48.89 \pm 26.67$ \\
 & \textcolor{red}{CSProp (ours)} &  $91.29 \pm 0.00$ & $98.49 \pm 0.00$ & $94.42 \pm 0.00$ & $97.36 \pm 0.00$ & $90.19 \pm 0.00$ & $78.33 \pm 0.00$ \\
 & \textbf{SimProp (ours)} &  $88.73 \pm 0.00$ & $86.30 \pm 0.00$ & $82.39 \pm 0.00$ & $80.87 \pm 0.00$ & $86.90 \pm 0.00$ & $91.59 \pm 0.00$ \\
\multirow{-8}{*}{KSC} & KNNProp (ours) & $80.42 \pm 0.00$ & $82.76 \pm 0.00$ & $74.89 \pm 0.00$ & $75.19 \pm 0.00$ & $77.38 \pm 0.00$ & $76.93 \pm 0.00$  \\ \hline
 & OSC-SCL \cite{osc-scl} & $80.92 \pm 1.47$ & $87.16 \pm 1.40$ & $87.97 \pm 1.09$ & $93.67 \pm 0.73$ & $78.85 \pm 1.61$ & $30.05 \pm 15.78$ \\
 & SSMLP-RPL \cite{ssmlp-rpl} &  $84.14 \pm 2.28$ & $85.67 \pm 2.67$ & $89.13 \pm 1.47$ & $90.82 \pm 1.38$ & $82.32 \pm 2.52$ & $74.76 \pm 11.29$ \\
 & HyperCASR \cite{hypercasr} & $79.47 \pm 1.49$ & $87.09 \pm 1.72$ & $86.70 \pm 1.38$ & $93.55 \pm 1.43$ & $77.20 \pm 1.64$ & $8.35 \pm 5.94$ \\
 & CACL \cite{cacl} & $89.87 \pm 1.73$ & $94.74 \pm 0.81$ & $92.67 \pm 1.36$ & $96.90 \pm 0.51$ & $88.69 \pm 1.92$ & $56.45 \pm 15.09$ \\
 & \textcolor{red}{CSProp (ours)} & $94.43 \pm 0.60$ & $96.39 \pm 0.57$ & $96.04 \pm 0.52$ & $97.78 \pm 0.46$ & $93.78 \pm 0.67$ & $84.41 \pm 1.74$  \\
 & SimProp (ours) & $86.03 \pm 1.36$ & $87.41 \pm 1.47$ & $88.73 \pm 1.06$ & $90.09 \pm 1.17$ & $84.42 \pm 1.50$ & $79.16 \pm 1.80$ \\
\multirow{-8}{*}{SA} & \textbf{KNNProp (ours)} & $92.98 \pm 1.33$ & $92.70 \pm 1.40$ & $94.66 \pm 0.77$ & $94.57 \pm 0.82$ & $92.16 \pm 1.47$ & $94.18 \pm 1.15$ \\ \hline
\end{tabular}%
}
}
\end{table*}

Regarding the overall performance and dataset-specific behaviors, the quantitative results presented in Tables \ref{tab:quantitative_5}, \ref{tab:quantitative_10}, and \ref{tab:quantitative_15} show the robust overall performance of the proposed FOSSeL framework across the varying few-shot scenarios. Our framework, with varying pixel prediction algorithms, consistently ranks among the best-performing methods. Notably, in all scenarios for SA and KSC, two different variants of our framework occupy the top two positions. On the IP dataset, our framework showed itself to be capable of achieving competitive results, securing the second-best performance in terms of H-Score in the 5-shot scenario and the best performance on the remaining ones. As previously mentioned, this decline in performance compared to the other two datasets can be inferred to be caused by a high intra-class variance for the dataset, which causes different clusters for the same class to be separated in the feature space, as seen in Figure \ref{fig:ip_features}. However, even in the 5-shot scenario, our framework achieved considerably better results than the other supervised baselines apart from CACL.

While the CACL \cite{cacl} baseline achieves better results in specific cases (most notably on the IP dataset), this comparison must be contextualized by the differences in the learning paradigms. It is important to note that our proposed FOSSeL method is trained entirely without supervision, and the final pixel prediction is done without any subsequent fine-tuning or training on the labeled support set. Consequently, our method operates at a natural disadvantage, as the experimental protocol did not allow us to adapt its learned feature space using the explicit knowledge provided by the labeled samples. The fact that FOSSeL achieves highly competitive (and often superior) results under these constraints highlights the representational power of the unsupervised feature space it generates.

Regarding OSC-SCL \cite{osc-scl} and SSMLP-RPL \cite{ssmlp-rpl} methods, the results show representational limits across varying data distributions. These two methods exhibit considerable dataset sensitivity: SSMLP-RPL achieves competitive results for the SA dataset, even surpassing CACL, but is still behind FOSSeL's variants, whereas the results on the KSC dataset demonstrate that the method can not maintain the same level of known/unknown class accuracy balance. OSC-SCL, on the other hand, presents the opposite characteristic by achieving a better balance for the KSC dataset, while not keeping the same performance on SA. HyperCASR consistently achieves the least balanced results, with the overall worst H-Score performance despite competitive results in terms of BA Open, while still not being able to surpass FOSSeL's performance. Without employing spatial-feature clustering techniques as our framework, these methods lack the representational power to fully exploit the limited available set of labeled pixels, reinforcing the necessity of our proposed spatial propagation strategies.

Considering our proposed methods, an interesting behavior is observed within the KSC dataset, where the performance of our variants remains similar between the 10-shot and 15-shot scenarios. This indicates that with just 10 labeled samples per class, our method acquires sufficient information to get close to its performance ceiling. This rapid saturation is likely due to the characteristics of the KSC dataset, which naturally facilitate the efficient exploitation of spatial information, allowing our propagation algorithms to converge early without needing additional labeled pixels. Also, an overall trend of diminishing returns is evident across all datasets. For our proposed methods, the improvement in prediction accuracy from the 5-shot to the 10-shot scenario is larger than the subsequent improvement from 10 to 15 shots. This shows that the FOSSeL framework requires only a few seeds (approximately 10 per class) to accurately map the topology of the classes, especially when the SSL training succeeds best.

Regarding the evaluation of the results in an open scenario versus closed scenarios, the results show FOSSeL's capacity to maintain accurate predictions for the known classes even when the prediction is opened to include unknown classes. As expected, closed-set metrics are higher than open-set metrics across most of the methods and datasets, as the open-set paradigm introduces important challenges to the prediction. However, our proposed methods maintain a close balance between the two, which can be observed due to the overall best performance according to the H-Score. That happens because FOSSeL relies on label propagation through an unsupervised feature space, naturally bounding the known classes based on their intrinsic clustering. Consequently, opening the predictions does not substantially degrade the accuracy of the known classes, which is extremely important when employing these methods in real-world settings.

Finally, the evaluation across the three datasets shows that no single label propagation variant is universally optimal. Instead, their performance depends on the scenario and the characteristics of the datasets:
\begin{itemize}
    \item CSProp enforces a direct, greedy assignment, according to the closest labeled seed. This characteristic is better for the correct classification of known categories, which can be seen by how it achieves the highest open balanced accuracy (BA Open) overall. This comes from a fundamental characteristic of CSProp, as it is a deterministic, parameter-free approach to open-set recognition. Rather than requiring an empirically tuned confidence threshold to separate known from unknown pixels, CSProp establishes a dynamic decision boundary based on the mean distance to the labeled seeds. While this mechanism simplifies the operational pipeline by removing a hyperparameter, it introduces a strict spatial assumption that can become a limitation depending on the underlying feature distribution. In contrast, the SimProp and KNNProp variants utilize explicitly defined confidence thresholds. This parameterized approach provides flexibility, allowing for granular control to balance the trade-off between known-class prioritization and unknown-class isolation.

    \item SimProp's row-stochastic transition matrix approach prevents overconfidence by ensuring that the propagation flow is probabilistically distributed. This creates the best balance between classifying known targets and rejecting unknown ones, which is empirically validated by SimProp achieving the highest H-Scores on the IP and KSC datasets. However, this global diffusion inherently dilutes the label probability at semantic boundaries and regions combining multiple classes. For a pixel to be classified as a known class in SimProp, its confidence must surpass an explicitly defined threshold. Thus, ambiguous pixels frequently fail to meet this threshold and are classified as unknown. This explains why SimProp usually can not classify known pixels as well as CSProp (which can be observed by comparing their closed metrics), but ultimately achieves the most balanced open-set detection.

    \item Conversely, KNNProp distinctively exploits the distribution of the self-supervised feature space by connecting vertices based on their proximity in the feature space, rather than relying exclusively on spatial pixel connections. The advantage of this localized propagation is that it directly exploits the intrinsic clustering quality of the SSL encoder. Consequently, KNNProp can be the best choice in scenarios where the dataset produces a highly separable and well-distributed feature space, such as the SA dataset (as seen in Figure~\ref{fig:sa_features}). In these scenarios, the nearest-neighbor constraint allows for the propagation to accurately occur inside the clusters without diluting label confidence across boundaries. However, this dependency also means that KNNProp's efficacy is strictly bound to the robustness of the SSL representation, as poorly separated feature spaces will cause misclassifications. This can be observed for the KSC dataset, where KNNProp achieves the overall worst results across the three proposed variants.
\end{itemize}

\subsubsection{Time Cost Analysis} 
To provide a comprehensive assessment of the framework's operational practicality, Table \ref{tab:time_cost} presents a quantitative comparison of the computational time costs for both the training and inference stages for the baselines and FOSSeL's variations. As observed in the table, the proposed framework requires a significantly higher training time (approximately 120,000 seconds) compared to baseline methods like HyperCASR or OSC-SCL. This disparity is a direct consequence of the architectural design. Our 120,000-second training duration represents an exhaustive self-supervised learning phase, completely independent of label annotation. While computationally intensive, this offline pre-training is a one-time cost that eliminates the need for dense human annotations to construct a robust feature space.

On the other hand, FOSSeL's inference time, which is the most critical metric for real-world operational deployment, is the lowest in all cases, except for SimProp in the KSC dataset. Once the feature space is established, the proposed CSProp method operates with exceptional speed. For instance, on the SA dataset, CSProp requires only 0.6 seconds for complete label propagation, vastly outperforming baselines like SSMLP-RPL (705 seconds) and OSC-SCL (64 seconds). A similar efficiency is observed on the Kennedy Space Center (KSC) dataset, where CSProp executes in just 1 second, compared to 47 seconds for CACL.

These results show that the proposed framework heavily front-loads its computational burden into the offline, self-supervised training phase, but it yields fast inference, which is highly desirable for rapid open-set pixel-wise classification upon deployment.

\begin{table}[h]
    \centering
    \caption{Computational time (in seconds) for training and inference stages averaged over different few-shot scenarios for the same dataset. The training times for the proposed methods encompass the full, unlabeled self-supervised phase.}
    \label{tab:time_cost}
    \begin{tabular}{llrr}
        \hline
        Dataset & Method    & \makecell{Training \\ Time (s)} & \makecell{Inference \\ Time (s)} \\
        \hline
        \multirow{7}{*}{IP}
        & OSC-SCL\cite{osc-scl}   & 259 & 12 \\
        & SSMLP-RPL\cite{ssmlp-rpl} & 483 & 75 \\
        & HyperCASR\cite{hypercasr} & 141  & 8 \\
        & CACL\cite{cacl}      & 4179 & 4 \\
        & CSProp    & $\approx120000$         & 0.1 \\
        & SimProp   & $\approx120000$         & 4 \\
        & KNNProp   & $\approx120000$         & 0.3 \\
        \hline
        \multirow{7}{*}{KSC}
        & OSC-SCL\cite{osc-scl}   & 176 & 6 \\
        & SSMLP-RPL\cite{ssmlp-rpl} & 336 & 36 \\
        & HyperCASR\cite{hypercasr} & 154   & 6 \\
        & CACL\cite{cacl}      & 3971 & 47 \\
        & CSProp    & $\approx120000$         & 1 \\
        & SimProp   & $\approx120000$         & 45 \\
        & KNNProp   & $\approx120000$         & 2 \\
        \hline
        \multirow{7}{*}{SA}
        & OSC-SCL\cite{osc-scl}   & 243 & 64 \\
        & SSMLP-RPL\cite{ssmlp-rpl} & 830 & 705 \\
        & HyperCASR\cite{hypercasr} & 178 & 66 \\
        & CACL\cite{cacl}      & 4824 & 17 \\
        & CSProp    & $\approx120000$         & 0.6 \\
        & SimProp   & $\approx120000$         & 8 \\
        & KNNProp   & $\approx120000$         & 0.8 \\
        \hline
    \end{tabular}
\end{table}

Despite its strong performance, the proposed framework presents some limitations. In particular, the reliance on superpixel segmentation introduces sensitivity to segmentation quality, which can affect performance when neighboring classes are grouped together, as explored in the following Section \ref{subsec:sensitivity}. Additionally, the open-set prediction relies on simple thresholding strategies, which, although effective in practice, could be further improved with adaptive or learned criteria. These aspects represent promising directions for future work.

\subsection{Analysis of Sensitivity to Hyperparameters} \label{subsec:sensitivity}

To evaluate the robustness and stability of the proposed framework, this subsection presents a sensitivity analysis of its two main hyperparameters previously introduced: the number of generated superpixels (number of SLIC segments, in this case) and the confidence score threshold $\tau$. To assess how these hyperparameter variations impact the trade-off between closed-set and unknown pixel accuracy, we utilize the $H$-score as our primary performance indicator.

\subsubsection{Number of segments} To evaluate the robustness of the FOSSeL framework regarding its spatial contextualization via superpixels, we conducted the sensitivity analysis with varying granularities: 500, 1000, and 2000 segments for the Indian Pines (IP) and Salinas (SA) datasets, and 2500, 5000, and 10000 segments for the Kennedy Space Center (KSC) dataset.

To isolate the impact of spatial granularity, we conducted the evaluation of the number of segments using a fixed, baseline confidence threshold of $0.5$ across all seed configurations for SimProp and KNNProp. While a $0.5$ threshold is well adapted for a highly constrained 5-shot scenario, it is less ideal when more labels are available (in this case, with 10 or 15 seeds). However, applying an optimized, highly restrictive threshold at this stage would act as an aggressive outlier filter, masking the underlying topological vulnerabilities caused by a sub-optimal segmentation. By maintaining a permissive baseline threshold that accepts the closed-set classification of low-certainty samples, we can better observe the impact that the number of segments imposes on the label propagation process.

\begin{figure*}[ht!]
\centering
\subfloat[]{\includegraphics[width=0.99\linewidth]{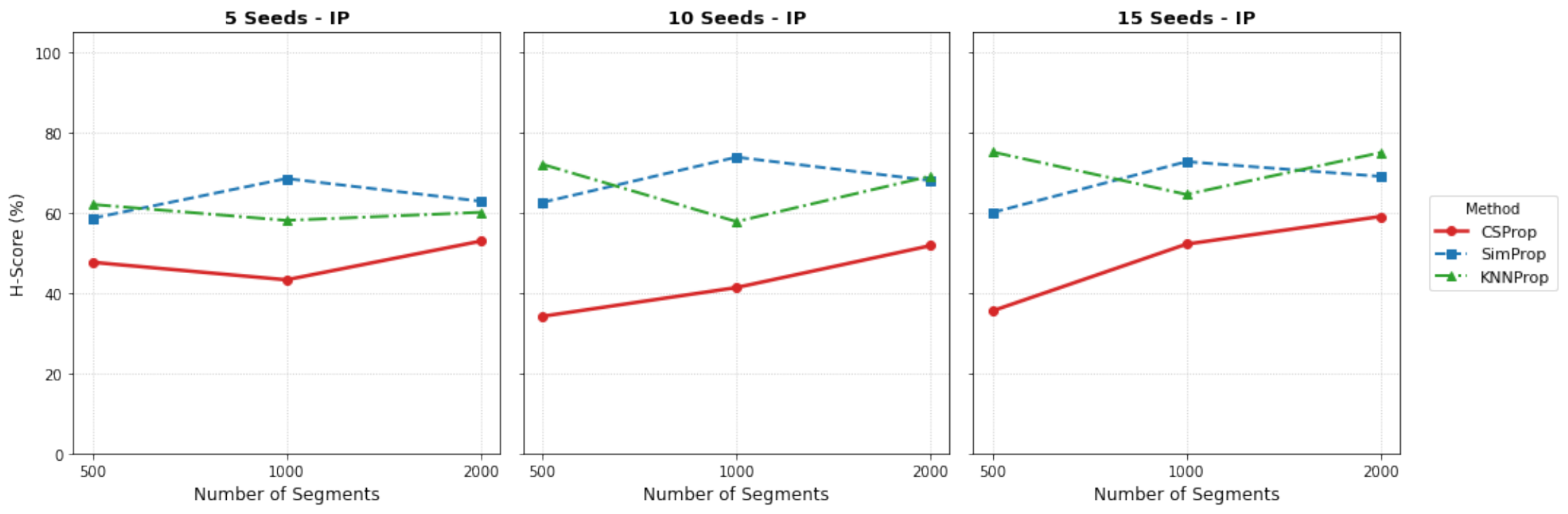}%
\label{fig:ip_segments}}
\\
\subfloat[]{\includegraphics[width=0.99\linewidth]{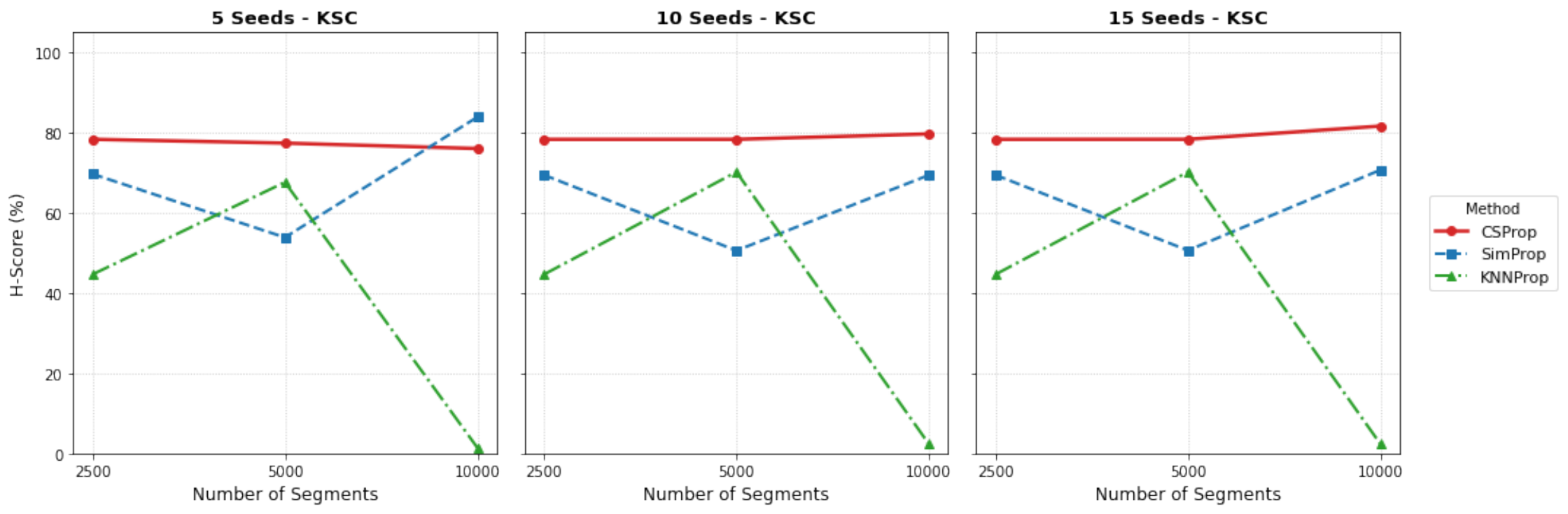}%
\label{fig:ksc_segments}}
\\
\subfloat[]{\includegraphics[width=0.99\linewidth]{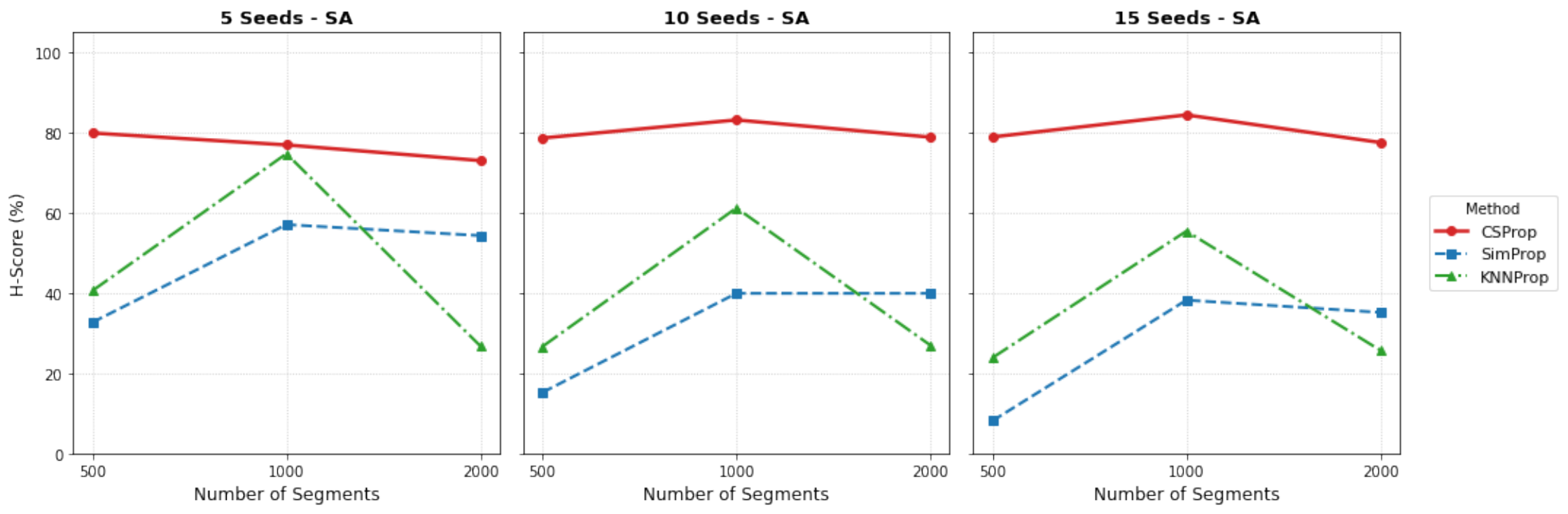}%
\label{fig:sa_segments}}
\caption{FOSSeL's performance in terms of H-Score according to variations on the number of generated superpixel segments and the number of available seeds. (a) Indian Pines. (b) Kennedy Space Center. (c) Salinas.}
\label{fig:segments_all}
\end{figure*}

The results of the sensitivity analysis regarding the number of segments can be seen in Figure \ref{fig:segments_all}. Table~\ref{tab:segments} complements these results by showing how many superpixels were created according to each experiment. The results demonstrate that increasing the number of superpixels does not universally improve performance. Rather, over-segmentation can lead to degradation of the H-score. In datasets characterized by lower spatial resolution, such as IP, using too few superpixels can limit the framework’s discriminative capacity. For example, at 500 segments in IP, the boundaries between classes become too coarse, frequently resulting in under-segmentations where distinct classes, including the unknown one, get combined within a single superpixel, negatively impacting the performance. Consequently, an intermediate granularity of 1000 segments emerges as the most stable configuration for IP across the different few-shot scenarios.

For the SA dataset, performance peaks at 1000 segments before degrading at higher granularities. This degradation exposes a fundamental vulnerability in graph-based label propagation related to over-segmentation: as the number of superpixels increases, the distinctions between neighboring regions become less apparent, effectively blurring the natural boundaries that separate different classes. During graph propagation, this boundary degradation causes labels to erroneously ``leak'' from known seeds into unwanted unknown regions with a confidence higher than the baseline threshold. This vulnerability can also be observed in the KSC dataset: while intermediate segmentations provide balanced results, an over-segmentation (such as with 10000 segments) severely impacts algorithms like KNNProp, since almost all unknown pixels are being misclassified as known classes due to the label leakage, dropping the $H$-score to near zero. With a higher confidence score of $0.7$, KNNProp's H-score results on KSC would increase to $55\%$ with 10 seeds, for example, which is considerably more balanced in terms of known and unknown classes' accuracy. These observations highlight one of the advantages of the proposed CSProp method, as it does not require additional hyperparameter tuning, remaining stable across varying scenarios.

Table \ref{tab:segments} details the relationship between the spatial granularity and the resulting node purity. As expected, increasing the number of generated segments boosts the percentage of pure superpixels, culminating in values close to or equal to 100\% purity in all datasets. However, correlating to the previous discussion, maximizing individual node purity through heavy over-segmentation can be detrimental. This happens because although these highly fragmented nodes are internally homogeneous, their reduced footprint blurs the transitions between distinct regions, allowing for labels to erroneously leak across the graph. Therefore, the empirical data validate that intermediate segmentations that secure sufficiently high purity (e.g., 90\% to 95\%) provide the optimal balance.

\begin{table}[h]
    \centering
    \caption{Number of generated superpixels (segments) and the percentage of pure superpixels, defined as those where all labeled pixels belong to a single known class. Superpixels consisting entirely of unlabeled background pixels are excluded from this calculation.}
    \label{tab:segments}
    \begin{tabular}{lrrr}
        \hline
        Dataset & SLIC Segments & Generated Segments & Pure Segments \% \\
        \hline
        \multirow{3}{*}{IP} 
            & 500  & 536  & 87 \\
            & 1000 & 793  & 90 \\
            & 2000 & 2247 & 97 \\
        \hline
        \multirow{3}{*}{KSC} 
            & 2500  & 2463 & 98 \\
            & 5000  & 4651 & 98 \\
            & 10000 & 8207 & 100 \\
        \hline
        \multirow{3}{*}{SA} 
            & 500  & 436  & 90 \\
            & 1000 & 869  & 95 \\
            & 2000 & 2190 & 98 \\
        \hline
    \end{tabular}
\end{table}

\subsubsection{Confidence Threshold}

\begin{figure*}[ht!]
\centering
\subfloat[]{\includegraphics[width=0.99\linewidth]{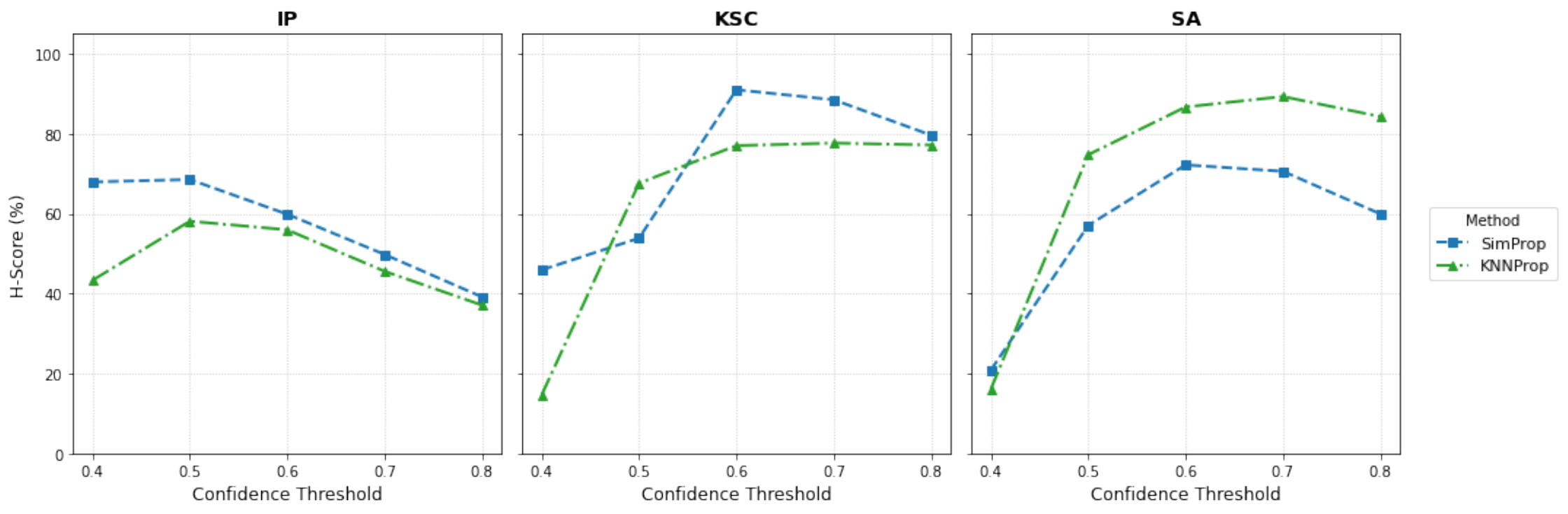}%
\label{fig:thresholds_5}}
\\
\subfloat[]{\includegraphics[width=0.99\linewidth]{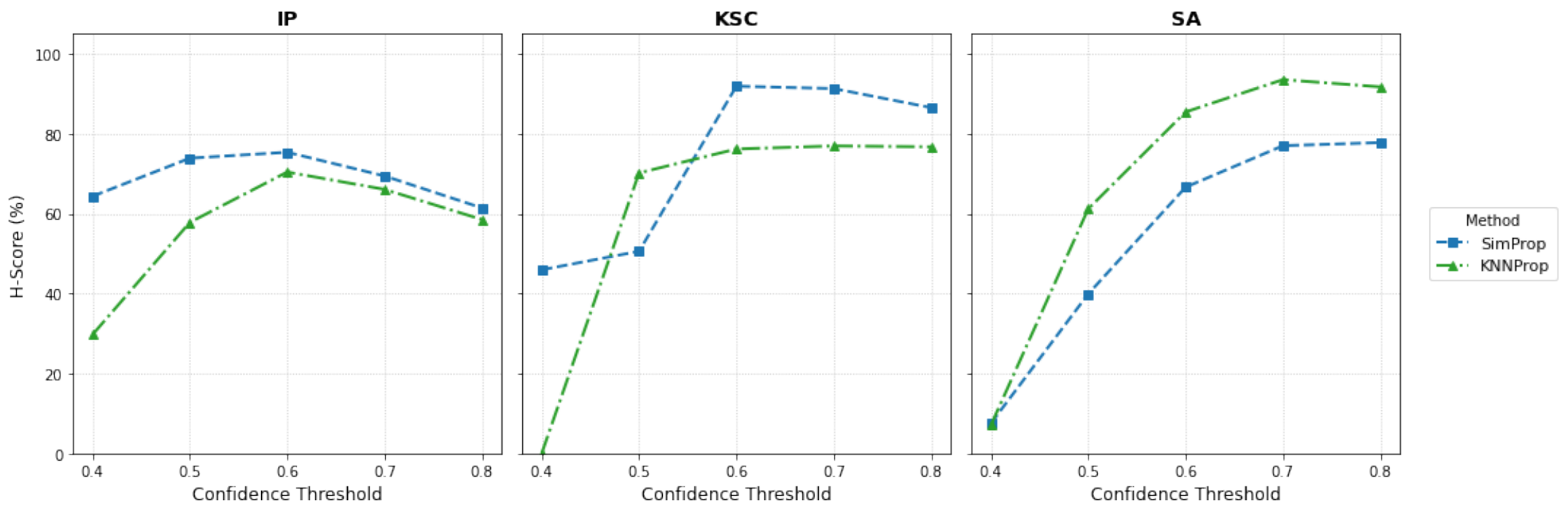}%
\label{fig:thresholds_10}}
\\
\subfloat[]{\includegraphics[width=0.99\linewidth]{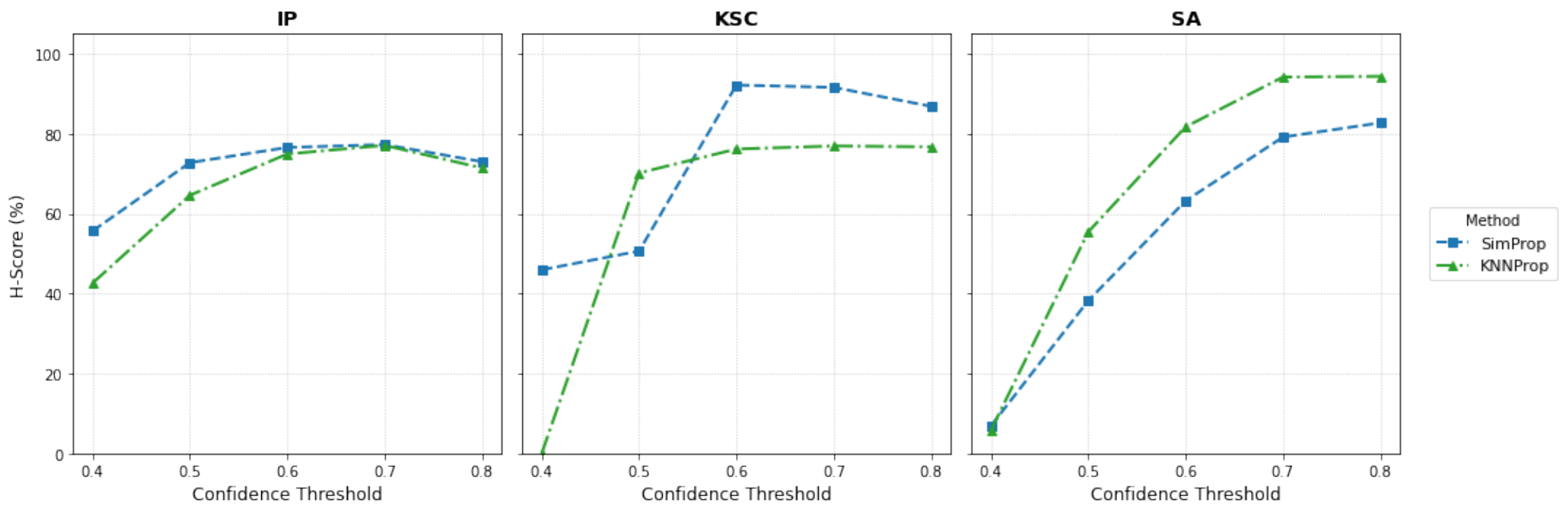}%
\label{fig:thresholds_15}}
\caption{Performance of SimProp and KNNProp, in terms of H-Score, as the confidence threshold $\tau$ varies from $0.4$ to $0.8$ for each few-shot scenario. (a) 5 seeds. (b) 10 seeds. (c) 15 seeds.}
\label{fig:thresholds_all}
\end{figure*}

Having established the spatial granularity for each dataset (1000 segments for IP and SA, and 5000 segments for KSC), the next phase of our evaluation investigates the impact of the confidence score threshold $\tau$ employed by the SimProp and KNNProp methods. This threshold dictates how strict the open-set rejection mechanism is. To assess how this parameter regulates the trade-off between classifying known classes and identifying unknown pixels, we varied the threshold from 0.4 to 0.8 across all three few-shot scenarios. The resulting H-scores are presented in Figure~\ref{fig:thresholds_all}.

As shown in the figure, overly permissive thresholds (e.g., 0.4) consistently lead to a severe performance degradation in open-set recognition. At these lower confidence levels, the model classifies almost all unknown pixels as known targets due to the low confidence needed to assign a label. This is particularly problematic for KNNProp, as it makes its H-score drop toward zero in multiple scenarios.

The results reveal that the optimal confidence threshold is dependent on the abundance of labeled data. In extreme few-shot scenarios (such as the 5-seed configuration), the overall confidence propagated through the graph is inherently lower. Thus, employing a threshold between 0.5 and 0.6 provides the optimal balance. Conversely, as the availability of labeled seeds increases in the 10- and 15-shot scenarios, the framework requires stricter thresholds, with the best performances consistently shifting toward the $\tau$ values of 0.6 and 0.7.

The data also demonstrates that pushing the threshold further to higher values (e.g., 0.8) makes it so that the known classes' predictions become too difficult, causing pixels from known classes to be erroneously classified as unknown. This explains the start of the decline in H-scores observed at the far right of the plots across nearly all configurations.

\subsection{Fairness of Baseline Comparison and Practical Implications}

An important aspect of our evaluation is the asymmetric nature of comparing FOSSeL, which relies entirely on an unsupervised feature space, against supervised baselines such as CACL\cite{cacl}, HyperCASR\cite{hypercasr}, SSMLP-RPL\cite{ssmlp-rpl} and OSC-SCL\cite{osc-scl}. We acknowledge that this comparison presents inherent fairness issues, as supervised methods possess the advantage of actively adapting their feature representations using the labeled support set. Conversely, FOSSeL operates at a natural disadvantage, functioning as a ``supervision-free'' classification framework, utilizing the few labeled seeds strictly for graph-based label propagation without any backpropagation or fine-tuning. However, we argue that this comparison serves to rigorously benchmark our framework. The empirical results demonstrate that FOSSeL frequently surpasses these supervised baselines in highly constrained few-shot scenarios (e.g., 5-shot). Furthermore, while the deployment of our method still requires a minimal set of annotated seeds for the final prediction phase, evaluating the capabilities of unsupervised methodologies holds significant scientific value. It proves that unannotated hyperspectral data can be effectively structured and clustered prior to any human annotation, serving as a robust foundation that may inspire future advancements in semi-supervised and zero-shot remote sensing research for open-set problems.

\section{Conclusion} \label{sec:conclusion}

In this paper, we propose FOSSeL, a framework that performs open-set hyperspectral image pixel-wise classification based on the feature space generated by a self-supervised learning method. We showed how SSL, in our case represented by the VICReg \cite{vicreg} method, can learn how to generate an adequate feature space for HSI without the need for labeled samples. We proposed three different pixel prediction algorithms that can be employed at the end of the FOSSeL framework, which shows how the framework is generalizable and allows for different approaches according to the characteristics of the problem, even without the need for further training or fine-tuning. Through the evaluation across three distinct hyperspectral datasets, we demonstrated that unsupervised feature learning, coupled with superpixel segmentation and label propagation from a few seeds, can effectively compete with, and frequently surpass, supervised baselines.

The pixel prediction methods explored in this work do not require many labeled samples, which is an important characteristic for dealing with HSI problems, known for being data-scarce. The framework exhibited a high efficiency in few-shot scenarios evaluated under the open-set paradigm, achieving a competitive performance with as few as 5 or 10 labeled samples per known class. FOSSeL also demonstrated statistical stability and robustness across diverse datasets without negatively impacting the accuracy of the known classes. However, we note that as the availability of labeled data increases, supervised baselines can eventually surpass our framework, especially on specific datasets where the self-supervised learning generates a less organized feature space.

The results indicate that combining self-supervised feature learning with spatially aware graph-based inference is a promising direction for addressing real-world hyperspectral classification challenges, particularly in scenarios with limited annotations and unknown classes. Because FOSSeL is designed as a highly modular framework, a promising avenue for future research involves a rigorous empirical benchmarking of various SSL architectures. Exploring how different self-supervised paradigms and their respective hyperparameter configurations influence the structure of the feature space will provide deeper insights into optimizing the subsequent open-set label propagation phase.

Finally, future work will explore how FOSSeL can be expanded to deal with domain adaptation problems, with the objective of using information from different data sources compared to the target dataset, such as from a different geographical location. We also plan to expand the use of FOSSeL to non-hyperspectral remote sensing data, which is a type of data with different characteristics that would require different techniques for the SSL training and pixel prediction methods. Allowing FOSSeL to work under these circumstances would amplify its use in multiple scenarios, such as cases in which the dataset presents a higher spatial resolution, and even datasets with more fine-grained details for the classes.

\ifCLASSOPTIONcaptionsoff
  \newpage
\fi

\bibliographystyle{IEEEtran}
\bibliography{bibtex/refs}

\begin{IEEEbiography}[{\includegraphics[width=1in,height=1.25in,clip,keepaspectratio]{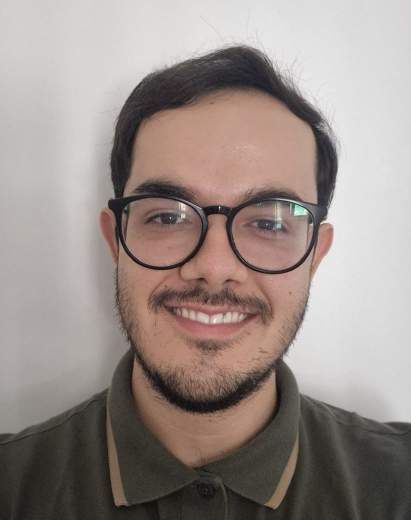}}]{Matheus B. Pereira}
is currently pursuing a joint Ph.D. in computer science at the Universidade Federal de Minas Gerais (UFMG), Brazil, and Université Gustave Eiffel, France. He also serves as an Assistant Professor at the Pontifical Catholic University of Minas Gerais. He received his bachelor’s degree in computer science from the Universidade Federal de Lavras (UFLA) and his master’s degree in computer science from UFMG. His research interests include computer vision, pattern recognition, and machine and deep learning, primarily applied to remote sensing data analysis.
\end{IEEEbiography}

\begin{IEEEbiography}[{\includegraphics[width=1in,height=1.25in,clip,keepaspectratio]{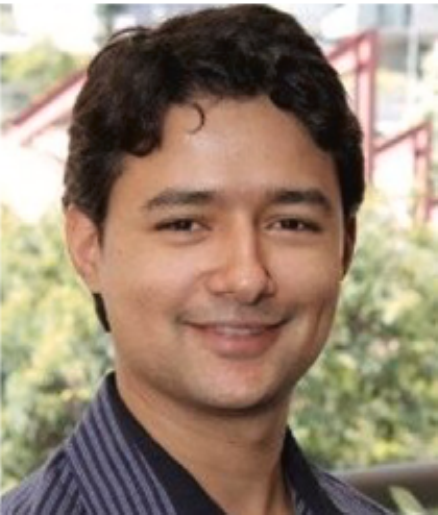}}]{Jefersson A. dos Santos}
(he/him) is currently an Assistant Professor (Lecturer) in Computer Vision at the University of Sheffield, United Kingdom. He used to hold a position as Associate Professor at the Universidade Federal de Minas Gerais, Brazil (2013–2022). Jefersson got a Ph.D. in Computer Science at the University of Campinas (Unicamp), Brazil, and at the University of Cergy-Pontoise, France, in 2013. He earned a Masters Degree in Computer Science from Unicamp in 2009 and a Bachelor Degree in Computer Science from University of Mato Grosso do Sul (UEMS), Brazil, in 2006. Jefersson is an IEEE Senior Member. He is currently an Associate Editor for the IEEE Geoscience and Remote Sensing Letters and Co-Chair for the ISPRS Working Group for AI/ML for Geospatial Data. Jefersson also serves as a PC member for several refereed journals and conferences and is currently the vice president of the IEEE GRSS Brazilian Chapter. Jefersson held the prestigious CNPq Productivity Research Scholarship (2016–2022). In 2021, he was awarded a competitive research grant from the Serrapilheira Institute, Brazil. His research interests include remote sensing image processing, computer vision and machine learning. Jefersson is the founder of the Laboratory of Pattern Recognition and Earth Observation (PATREO) at the Department of Computer Science, UFMG, Brazil.
\end{IEEEbiography}

\begin{IEEEbiography}[{\includegraphics[width=1in,height=1.25in,clip,keepaspectratio]{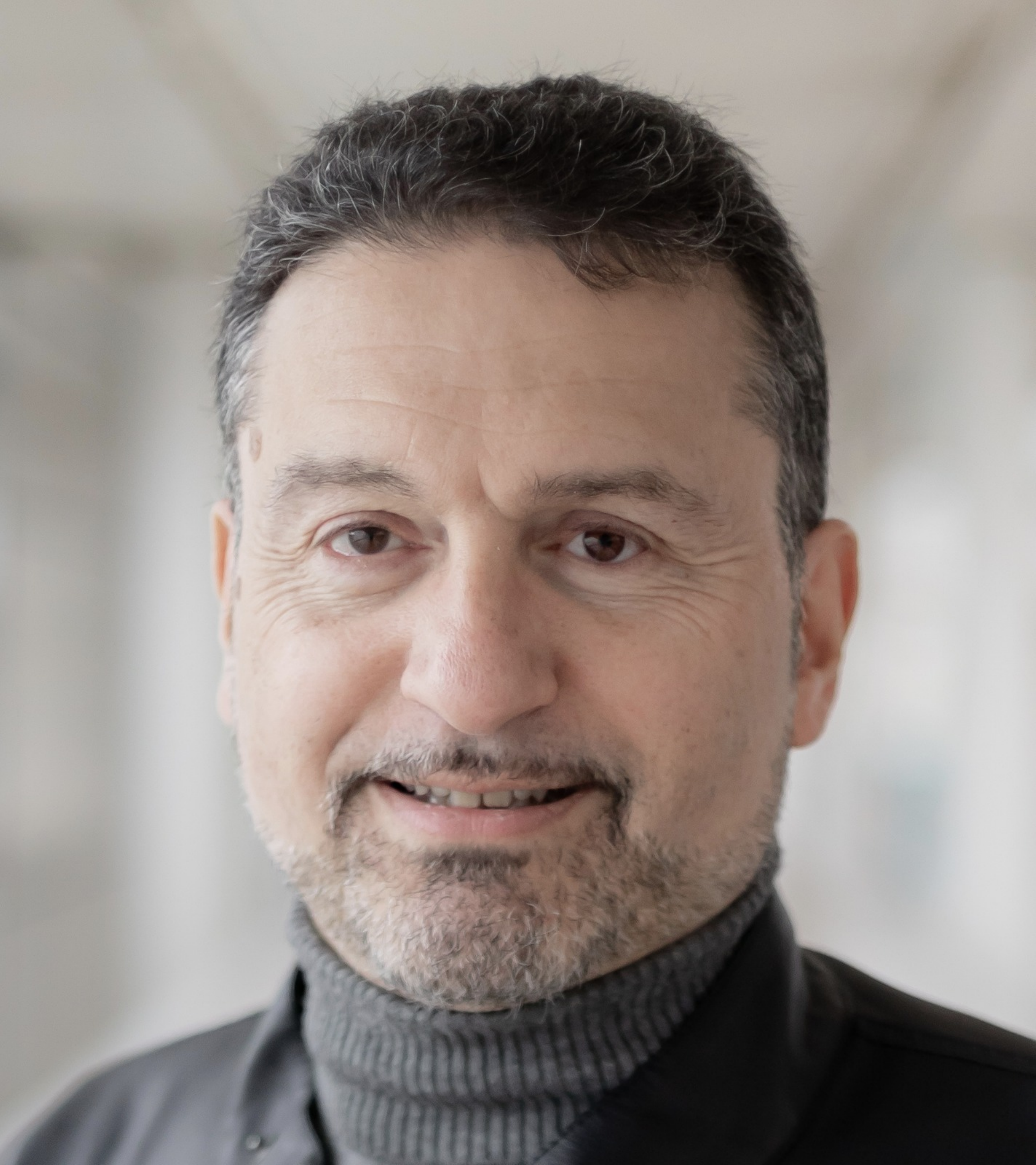}}]{Laurent Najman}
received the Habilitation á Diriger les Recherches in 2006 from University of Marne-la-Vallée, a Ph.D. of applied mathematics from Paris-Dauphine University in 1994 with the highest honor (Félicitations du Jury) and an ``Ingénieur'' degree from the Ecole des Mines de Paris in 1991. After earning his engineering degree, he worked in the central research laboratories of Thomson-CSF for three years, working on some problems of infrared image segmentation using mathematical morphology. He then joined a start-up company named Animation Science in 1995, as director of research and development. The technology of particle systems for computer graphics and scientific visualization, developed by the company under his technical leadership received several awards, including the ``European Information Technology Prize 1997'' awarded by the European Commission (Esprit programme) and by the European Council for Applied Science and Engineering and the ``Hottest Products of the Year 1996'' awarded by the Computer Graphics World journal. In 1998, he joined OCÉ Print Logic Technologies, as senior scientist. He worked there on various problems of image analysis dedicated to scanning and printing. In 2002, he joined the Computer Science Department of ESIEE Paris, where he is a Professor and a member of the Laboratoire Informatique Gaspard Monge (LIGM), Université Gustave Eiffel -- France. Since August 2025, he has been a Full Professor in the Department of Mathematics at Khalifa University -- UAE, where he is currently on leave from the LIGM. His current research interest is discrete mathematical morphology and discrete optimization.
\end{IEEEbiography}

\end{document}